\documentclass[fleqn,usenatbib]{mnras}

\usepackage{newtxtext,newtxmath}

\usepackage[T1]{fontenc}
\usepackage{ae,aecompl}

\usepackage{graphicx}	
\usepackage{amsmath}	
\usepackage{amssymb}	

\usepackage{comment}
\usepackage{xspace}     
\usepackage{xcolor}     

\usepackage{blindtext}

\newcommand{\xGASS}{xGASS\xspace}	
\newcommand{\ProFit}{\textsc{ProFit}\xspace}	
\newcommand{\ProFound}{\textsc{ProFound}\xspace}	
\newcommand{\GALEX}{\textit{GALEX}\xspace}	
\newcommand{\SDSS}{SDSS\xspace}	

\newcommand{\Mstar}{$M_{\star}$\xspace}    
\newcommand{\MstarT}{$M_{\star,Total}$\xspace}    
\newcommand{\MstarD}{$M_{\star,Disc}$\xspace}    

\newcommand{\SFR}{$SFR$\xspace}    
\newcommand{\sSFR}{$sSFR$\xspace}    

\newcommand{\HI}{H$\,\textsc{I}$\xspace}    
\newcommand{\Htwo}{H$_\mathrm{2}$\xspace}    
\newcommand{\MHI}{$M_\mathrm{HI}$\xspace}      

\newcommand{\gband}{$g$-band\xspace}            
\newcommand{\rband}{$r$-band\xspace}            
\newcommand{\iband}{$i$-band\xspace}            

\newcommand{\gi}{$g-i$\xspace}          
\newcommand{\NUVr}{NUV$\,-\,r$\xspace}              

\newcommand{\BtoTr}{$\beta_\mathrm{r}$\xspace}          
\newcommand{\BtoTi}{$\beta_\mathrm{i}$\xspace}          
\newcommand{\BtoTM}{$\beta_\mathrm{M_{\star}}$\xspace}          

\newcommand{\Msun}{$M_{\odot}$\xspace}    

\newcommand{\MHIvM}{$M_\mathrm{HI}$--$M_\mathrm{\star}$\xspace} 

\newcommand{\Sersic}{S\'{e}rsic\xspace}    

\newcommand{\simperc}[1] {$\sim#1$\,\%}

\title[The Impact of Bulges in HI Scaling Relations]{xGASS: The Impact of Photometric Bulges on the Scatter of HI Scaling Relations}

\author[R. H. W. Cook et al.]{
Robin H. W. Cook$^{1,2}$\thanks{E-mail: robin.cook@icrar.org},
Luca Cortese$^{1,2}$,
Barbara Catinella$^{1,2}$,
Aaron Robotham$^{1,2}$
\\
$^{1}$International Centre for Radio Astronomy Research (ICRAR), University of Western Australia, Crawley, WA 6009, Australia\\
$^{2}$Australian Research Council, Centre of Excellence for All Sky Astrophysics in 3 Dimensions (ASTRO 3D), Australia
}

\date{Accepted 2019 September 18. Received 2019 September 18; in original form 2019 April 26}

\pubyear{2019}

\begin{document}
\label{firstpage}
\pagerange{\pageref{firstpage}--\pageref{lastpage}}
\maketitle

\begin{abstract}
We present a structural decomposition analysis of the galaxies in the extended \GALEX Arecibo \SDSS Survey (xGASS) using (\textit{gri}) images from the Sloan Digital Sky Survey. Utilising the 2D Bayesian light profile fitting code ProFit, we fit single- and double-component models taking advantage of a robust Markov chain Monte Carlo optimisation algorithm in which we assume a \Sersic profile for single-component models and a combination of a \Sersic bulge and \emph{near}-exponential disc ($0.5 \leqslant n \leqslant 1.5$) for double-component models. We investigate the effect of bulges on the atomic hydrogen (\HI) content in galaxies by revisiting the \HI-to-stellar mass scaling relations with the bulge-to-total ratio measured in the ProFit decompositions. We show that, at both fixed total and disc stellar mass, more bulge-dominated galaxies have systematically lower \HI masses, implying that bulge-dominated galaxies with large \HI reservoirs are rare in the local Universe. We see similar trends when separating galaxies by a bulge-to-total ratio based either on luminosity or stellar mass, however, the trends are more evident with luminosity. Importantly, when controlling for both stellar mass and star formation rate, the separation of atomic gas content reduces to within 0.3 dex between galaxies of different bulge-to-total ratios. Our findings suggest that the presence of a photometric bulge has little effect on the global \HI gas reservoirs of local galaxies.
\end{abstract}




\section{Introduction}
\label{sec:introduction}
Galaxies in the local Universe can be broadly split into two distinct categories: star-forming and quiescent galaxies \citep{Roberts+Haynes1994,Blanton+Moustakas2009,Pozzetti2010}. The former are those where their current star formation rate appears closely associated to the accumulated stellar mass of the galaxy: forming the well-known star-forming main-sequence (SFMS). Here, the fractional growth of stellar mass (i.e. specific star formation rate) is only weakly dependent on the total stellar mass of a galaxy and has been shown to be in place since at least $z \sim 2$ \citep{Brinchmann2004,Daddi2007,Elbaz2007,Noeske2007,Salim2007}. The latter instead have significantly suppressed star formation rates and are distributed well below the main-sequence by $\sim1.5$\,--\,$2$ dex, or more \citep{Ilbert2010}. It is believed that the population of galaxies existing on the main-sequence will eventually evolve into a passive population through the shutting down of their star formation \citep{Bell2004}. Understanding the physical process(es) responsible for causing a galaxy to cease its evolution along the main-sequence (so-called ``quenching'') has long been a challenge in studies of galaxy evolution.

The emergence of large spectroscopic surveys of galaxies in the nearby Universe, such as the Sloan Digital Sky Survey (\SDSS; \citealt{York2000}), has prompted considerable effort into quantifying relationships seen between stellar mass, size, internal structure, star formation rate, metallicity and dust content \citep{Kauffmann2003, Brinchmann2004, Tremonti2004, Baldry2004, Baldry2006}. The observables that appear to most clearly accompany the shutting down of star formation in galaxies are those associated with galaxy morphology \citep{Wuyts2011}. Both simulations and observations have shown a strong link with structural quantities such as the concentration index $R_{90}/R_{50}$ \citep{Strateva2001,Catinella2010,Catinella2012}, stellar mass surface density $\mu_{\star}$ \citep{Kauffmann2003,Fang2013,Catinella2018}, \Sersic index $n$ \citep{Blanton2003,Schiminovich2007,Kelvin2012,Lange2015}, bulge-to-total ratio \citep{Kauffmann2012,Omand2014}, central velocity dispersion $\sigma$ \citep{Franx2008,Smith2009,Bell2012,Wake2012} as well as the inner 1\,kpc stellar mass surface density $\Sigma_{1\rm{kpc}}$ \citep{Cheung2012,Woo2015}. However, it has yet to be shown that this observed correlation is due to a direct causal link.

There are also complex relationships observed between the cold gas content and global galaxy properties such as the size, shape and star formation properties (see e.g. \citealt{Kennicutt1998} and references therein). However, it is still unclear how --- and to what extent --- the cold gas content of galaxies relates to these other fundamental properties; of particular importance is the interplay between galaxy morphology and cold gas reservoirs. This is in part due to the fact that the numbers of galaxies observed in cold gas surveys is lagging far behind those of optical surveys. In addition, as atomic hydrogen (\HI) is, on average, more difficult to detect in early-type galaxies, the samples studied are generally biased towards galaxies that are gas-rich which tend to be star-forming and late-type morphologies. Large area, blind \HI surveys such as the \HI Parkes All-Sky Survey (HIPASS; \citealt{Barnes2001,Meyer2004}) or the Arecibo Legacy Fast ALFA (ALFALFA; \citealt{Giovanelli2005,Haynes2018}) have proven invaluable as they provide uniform coverage over a large region of sky. However, such surveys are inherently shallow and are often not able to reach sufficient sensitivities to detect a significant population of gas-poor galaxies. In this study, we make use of \HI observations from the extended GALEX Arecibo SDSS Survey (xGASS; \citealt{Catinella2010,Catinella2018}) as this provides us with a statistically significant sample of cold gas observations that covers a representative range of morphological types, as well as probing gas-poor systems.

The apparent correlation between the star-forming properties of galaxies and their overall structure has often pointed to the conclusion that the internal processes in galaxies must be responsible for the quenching of galaxies. This could be considered an indirect effect where the structure of galaxies is assumed to be connected to the properties of their central black hole. It has been proposed that accretion onto a black hole is capable of removing cold gas via powerful quasar-driven winds \citep{Springel2005,Hopkins2006} in addition to causing the gas in the surrounding halo to be heated \citep{Croton2006}. Alternatively, this correlation could be seen as being directly related in a scenario where star formation is heavily influenced by the stability of cold gas in the disc \citep{Martig2009,Genzel2014}. The latter has motivated the ``morphological quenching'' scenario \citep{Martig2009} in which a gaseous disc can become stable against fragmentation to bound clumps in the presence of a large stellar spheroid. Recent observations from the ATLAS$^\mathrm{3D}$ Survey \citep{Cappellari2011} have suggested that early-type galaxies have star formation efficiencies ($SFE = SFR/M_{gas}$) that are a factor of $\sim2\times$ lower on the Kennicutt-Schmidt law than late-type galaxies \citep{Martig2013,Davis2014}. However, it is not yet clear whether this correlation between the quenching of star formation and morphology is due to a causality or the result of some other underlying physics (e.g. \citealt{Lilly2016}). Conversely, other authors \citep{Schiminovich2010,Schawinski2014} have shown that generally the opposite is true or that structure has no impact on the SFE. Interestingly, \citet{Fabello2011} showed that, at fixed $M_{\star}$ and \SFR, galaxies with higher concentration indices $R_{90}/R_{50}$ (indicative of more bulge-dominated systems) tended to contain smaller amounts of \HI gas than those with lower concentration indices.

It should be noted that those works supporting the morphological quenching scenario often compared Hubble classifications in which early-types would be usually taken from a passive population against late-types selected usually from a star-forming population. \citet{Brown2015} showed clearly that the scaling relations of gas fraction with stellar mass surface density are primarily driven by the underlying bimodality in the star formation properties of galaxies. This highlights the need for a sample of galaxies that is not biased towards gas-rich star-forming galaxies or selected via visual classification schemes.

In this paper, we perform a structural decomposition analysis on the light profiles of galaxies in the \xGASS sample to investigate whether the correlation observed between galaxy morphology and the quenching of star formation can be causally linked through an imprint on galaxy atomic gas reservoirs.
The paper is organised as follows. Section \ref{sec:xGASS} gives an overview of the \xGASS sample, Section \ref{sec:structural_decomposition} then describes the procedure taken to perform the structural decomposition of this sample and a validation of the measured model parameters in provided in Section \ref{sec:model_validation}. Section \ref{sec:results} presents an application of this structural information towards investigating the impact of bulges in maintaining the reservoirs of cold gas in galaxies. We discuss the implications of this work in Section \ref{sec:discussion} and conclude in Section \ref{sec:conclusion}.

Throughout the paper we use AB magnitudes and assume a \citet{Chabrier2003} initial mass function. All distance-dependent quantities are computed assuming $\Omega_{M} = 0.3$, $\Omega_{\Lambda} = 0.7$ and $H_{0} = 70$\,km\,s$^{-1}$\,Mpc$^{-1}$.

\section{The xGASS Sample}
\label{sec:xGASS}
The extended \GALEX Arecibo \SDSS Survey (xGASS; \citealt{Catinella2010,Catinella2018}) is a census of atomic hydrogen (\HI) gas content for 1179 galaxies selected only by redshift ($0.01 < z < 0.05$) and stellar mass ($M_{\star} = 10^{9}$\,--\,$10^{11.5}$\,\Msun). Unlike most previous \HI surveys, \xGASS is gas fraction limited in that the integration times of the Arecibo telescope observations were dictated by the requisite that they either be detected or reach a gas fraction limit of $M_\mathrm{H\textsc{I}}/M_{\star} \sim\;$2\,--\,10\% (dependent on \Mstar). In total, $\sim$1,300 hours of Arecibo telescope time was required which has led to this survey having the deepest observations of cold gas for galaxies in the local Universe to date. The sample provides a statistically significant number of galaxies with \HI observations over a wide range of stellar masses and morphologies. It is for these reasons that the \xGASS sample is currently the best suited for understanding the impact of bulges on the \HI content of galaxies. In this work, we focus only on the 1179 galaxies in the \xGASS representative sample and for the scientific analysis we have excluded the 74 \HI confused sources.

\xGASS galaxies have been selected from a parent sample in \SDSS DR6 \citep{Adelman-McCarthy2008} for which \SDSS spectroscopy and \textit{Galaxy Evolution Explorer} (\GALEX) observations were available and randomly sampled such that the resulting catalogue has a near flat distribution in stellar mass (see \citealt{Catinella2018} for more details). Optical parameters were obtained from the \SDSS DR7 \citep{Abazajian2009} database server whilst the UV photometry and star formation rates (SFR) are determined from the combination of near-UV (NUV) magnitudes (from various GALEX catalogues) with mid-infrared (MIR) from the Wide-field Infrared Survey Explorer (WISE, \citealt{Wright2010}) as detailed in \citet{Janowiecki2017}. Wherever there was no unflagged fluxes available in both NUV and MIR, the SFRs were determined from the spectroscopic energy distribution fits in \citet{Wang2011}.

\section{Structural Decomposition of Galaxies}
\label{sec:structural_decomposition}

In this section, we describe our usage of the two-dimensional structural decomposition code \ProFit \footnote{\href{https://cran.r-project.org/web/packages/ProFit/index.html}{https://cran.r-project.org/web/packages/ProFit/index.html}} \citep{Robotham2017} that has been designed to robustly decompose galaxy images into their separate bulge and disc components.

\subsection{PROFIT: Bayesian Profile Fitting}
\label{sec:ProFit}
At its core, \ProFit is a tool for Bayesian two-dimensional modelling of galaxy photometric profiles consisting of both a low-level \textsc{C}++ library for efficient pixel integration overlaid with a high-level \textsc{R} interface which facilitates likelihood calculations as well as various optimisation and sampling algorithms. Having access to the Central R Archive Network (\textsc{CRAN}) grants straightforward implementation of the wealth of statistical packages and optimisation methods (maximum-likelihood and Bayesian) developed originally in the context of advanced data analysis. This agnostic implementation of optimising routines permits the flexibility required to robustly fit galaxy light profiles with automated Bayesian modelling. This has been highlighted as one of the key reasons why previous structural decomposition studies have had a high rate of incorrectly modelled galaxies. For example, \citep{Lange2016} state that typically 20\,--\,30\% of automatic fits using Levenberg--Marquardt algorithms are, in some way, non-physical.

\subsection{Required Inputs}
\label{sec:required_inputs}

\begin{figure*}
    \centering
    \includegraphics[width=\textwidth]{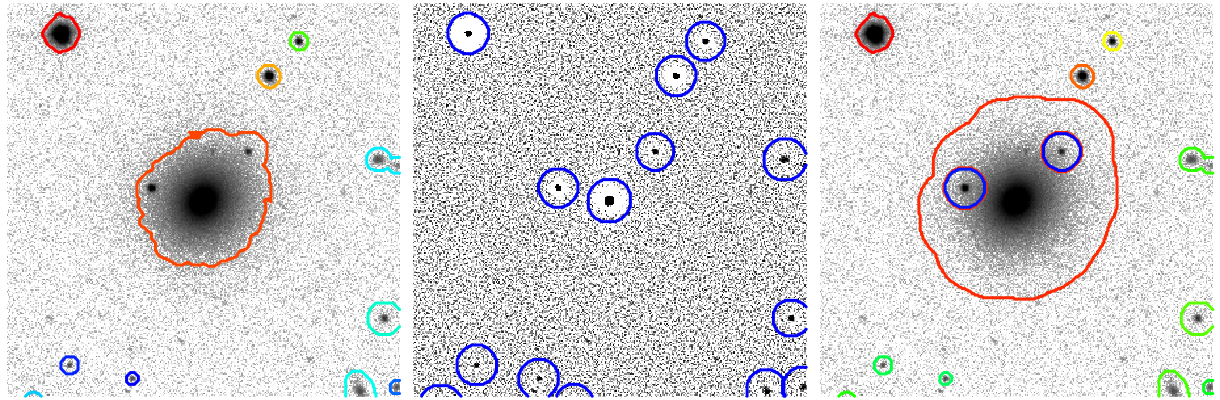}
    \caption{Stages of the segmenting process used for complex systems where pronounced spiral-arm structure or embedded sources result in a poor segmentation map. The left panel shows the initial source finding and merging of segments adjacent to the central segment. The middle panel shows the point-like sources detected within the residual of the blurred and original image. The right panel shows the final segmentation map with the red contour defining the galaxy segment and the remaining colours showing the unassociated objects; in particular, the embedded point sources that were masked (shown in blue).}
   \label{fig:segmentation}
\end{figure*}

A summary of the required observational data necessary to obtain a meaningful profile fit is given in \citet{Robotham2017}; here we describe the necessary inputs with relation to the data used in this study.\\

    \textbf{Photometric Images:} We have used \SDSS \rband images as the starting point for our models as these provide the best signal-to-noise ratio at a wavelength sensitive enough to the distinct structural components of galaxies. The moderate depth and spatial resolution of these data is suitable for the structural decomposition of \xGASS galaxies as it is a relatively nearby sample. The derived model parameters then inform upon subsequent models in the \gband and \iband images. The 95\% completeness limits for point sources are $g^{*} \leqslant 22.2$, $r^{*} \leqslant 22.2$ and $i^{*} \leqslant 21.3$ mag \citep{Stoughton2002}. All of the reduced frames (\textit{drC}) have been downloaded from the \SDSS Data Release 7 (DR7) and have previously been bias subtracted and flat-field corrected through the standard \SDSS \textit{frames} pipeline \citep{Stoughton2002}. The point-spread functions (PSF) for each frame have also been sourced from \SDSS DR7 through reconstruction of the corresponding \textit{psField} files. We make use of the \SDSS \emph{photo} software\footnote{https://www.sdss.org/dr12/software/} which fits a PSF separately in each \SDSS band as a function of position within an SDSS frame. The median seeing of all SDSS imaging data for the $g$-, $r$- and \iband are $1.44''$, $1.32''$ and $1.26''$, respectively.\\


We construct $\sim$\,600\,$\times$\,600 pixel ($4'\times4'$) cutouts centred upon each galaxy (with exception to the largest few galaxies) which accommodates the range of angular sizes of our sample whilst balancing the need for a suitable number of \textit{true} sky pixels to estimate the properties of background sky. Wherever a galaxy spans across the edge of two adjacent frames, we make no attempt to stitch the images together in an effort to improve the signal-to-noise as there are often complications in consolidating the differing PSFs and background sky statistics. In reality, one of the overlapping frames will contain both the necessary central region of the galaxy as well as sufficient light on at least one side of the outer regions to be properly modelled.

\textbf{Segmentation Maps:} The projection of any astronomical observation onto the two-dimensional plane of the sky collapses everything the telescope observes along that particular line of sight. For this reason, a \emph{segmentation map} is necessary to distinguish between the region(s) associated with each target galaxy as well as the region(s) of foreground stars and unassociated galaxies embedded within or near the target galaxy. We have used the \ProFound package\footnote{\href{https://cran.rstudio.com/web/packages/ProFound/index.html}{https://cran.rstudio.com/web/packages/ProFound/index.html}} \citep{Robotham2018} to perform an automated source extraction on the cutout images. \ProFound is based on a common digital ``Watershed algorithm'' for dividing images into distinct \emph{catchment regions} and has proven to be very powerful in separating distinct sources in astronomical data. Our image segmentation process begins with an initial identification of all sources present within an \rband image followed by a dilation of the target galaxy's segmentation region such that it sufficiently encompasses the surrounding sky pixels which act as an ``anchor'' to the background sky pedestal. See Appendix \ref{appendix:profound_parameters} for an outline of the basic parameters used when running \ProFound on our image data.

Each galaxy segmentation map was visually inspected to ensure that our model fits were not affected by poorly defined regions. In any automated source finding algorithm, a poor segmentation map can result for a number of reasons. The most common problems are a result of the fine balance between the fragmentation of flocculant galaxies against the combining of regions from superimposed or interacting objects and embedded bright stars. Using many of the convenience functions within the \ProFound package, we make an effort to repair poor segmentation maps by stitching together associated segments and/or masking out embedded sources that could otherwise influence the resulting light profile.

Figure \ref{fig:segmentation} shows an example of a complex segmentation map where both stitching and masking is required. The first stage involves the initial source-finding of individual segments from which those adjoining the central segment have been merged together and dilated to become the target segment. The second panel shows a difference map between the original image and one that has been blurred. By subtracting these images, extended structure will remain in both images and hence not be present in the residual, whereas sources with a steep light profile (e.g. stars and, for that matter, also a central bulge) will be pronounced. We use the measured properties (ellipticity, concentration, orientation, etc.) of the segments detected in the residual image to identify embedded sources that require masking. The final panel shows the merging of the main segment with the point source mask as well as a final dilation which extends the segment sufficiently far into the background sky.\\

\textbf{Background Sky Models:} In addition to the segmentation map, we use \ProFound to make an estimate of the local sky background which is then subtracted from the image. The background sky map is obtained by iteratively sigma clipping over the sky pixels defined from the segmentation map (until convergence is achieved) then linearly interpolating across a grid of size $1'\times1'$. For the few most extended galaxies in our sample, we required a slightly larger cutout image (and grid size) to include sufficient sky pixels for a valid background estimation.\\

\textbf{Error Maps:} The final data product that we extract from \ProFound is the sigma map or, its inverse, the weight map. In most astronomical images, individual pixels do not necessarily share the same statistical properties; differences can occur due to variable survey depth and noise properties, gain corrections, deviations in the readout of pixel counts, etc. The relative errors associated with these processes as well as the statistical uncertainty of measuring the photon count (i.e. ``shot noise'') in each pixel are combined to construct the weight map. This gives a prediction of the measurement precision at each pixel of the corresponding science image. We use this in the fitting process to appropriately down-weight the importance of pixels with a greater uncertainty in their measurement.

\subsection{Galaxy Light Profile Fitting}
\label{sec:galaxy_light_profiles}
The earliest attempts at characterising galaxy light profiles came from the use of the de Vaucouleurs' profile \citep{deVaucouleurs1948} to describe elliptical galaxies and, later, the discovery that galactic discs were better described by an exponentially declining profile \citep{deVaucouleurs1959}. It was, however, the generalisation of de Vaucouleurs' $R^{1/4}$ profile to the $R^{1/n}$ law by \citet{Sersic1963} that quickly became the de facto standard for fitting the wide range of shapes seen in dynamically hot stellar systems \citep{Graham+Guzman2003}. The \Sersic law is summarised by an equation that describes the intensity of light $I$ as a function of the radius $R$, given by

\begin{equation}
    I(R) = I_{e}\exp\left\{\,-b_{n}\,\left[\,\left(\,\frac{R}{R_{e}}\,\right)^{1/n}-1\right]\,\right\},
\end{equation}

\noindent
where $I_{e}$ is the intensity at the effective radius $R_{e}$ (the radius enclosing half of the total flux) and $n$ denotes the \Sersic index (or profile \emph{shape}) which has the notable cases of an exponential with $n = 1$, a de Vaucouleurs profile with $n = 4$ and a Gaussian with $n = 0.5$. The $b_{n}$ term is a derived quantity that ensures the correct integration properties at the effective radius. For a review of the behaviour of the profile and its quantities, see \citet{Graham+Driver2005} and references therein.

As has often been the case in comprehensive structural decomposition studies of well-resolved galaxies \citep{Anderdakis1995,Baggett1998,Khosroshahi2000,Graham2001,Allen2006,Gadotti2009,Simard2011,Meert2015,Kim2016,Lange2016,Fischer2018}, we have chosen to fit galaxies with two models: (1) a single \Sersic model with free \Sersic index $n$ and (2) a combination of a near-exponential disc (0.5 < $n$ < 1.5) plus a \Sersic bulge component ($n$ free). Although the underlying radial light profile of a rotationally-supported disc is expected to follow an exponentially-declining profile ($n = 1$), in reality, the effects of projection, asymmetries due to interactions or the otherwise diffuse structure of outer spiral arms can cause deviations away from a true exponential profile \citep{Pohlen2002,Erwin2005}. This choice is further substantiated by the distribution of \Sersic indices in pure-disc models which peaks at $n=1$ but also shows a significant spread from $n \simeq 0.5$ to $n \simeq 1.5$. \citet{Pohlen+Trujillo2006} showed that using imaging data from the SDSS survey, in a sample of $\sim$90 nearby spiral galaxies, only $\sim$10\% showed a purely exponential disc down to the noise level of the data. The remaining showed a break followed either by a down-bending to a steeper outer region ($\sim$60\%) or an up-bending to a shallower outer region ($\sim$30\%). Previous structural decomposition studies have highlighted the importance of accounting for truncated and anti-truncated discs, especially in well-resolved, deep sensitivity data \citep{Kim2014, Gao+Ho2017}. We use the flexibility in the disc \Sersic index to better accommodate this observed range of disc stellar light profiles. However, in some cases, it is clear that a fixed exponential disc profile performs better than if allowed to be free. We describe the selection of best-fit models in Section \ref{sec:model_filtering}.

\begin{figure*}
    \centering
    \includegraphics[width=0.625\textwidth]{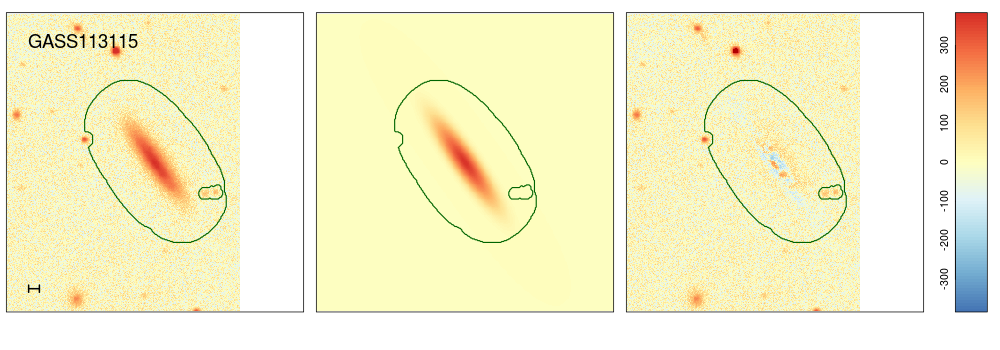}
    \includegraphics[width=0.325\textwidth]{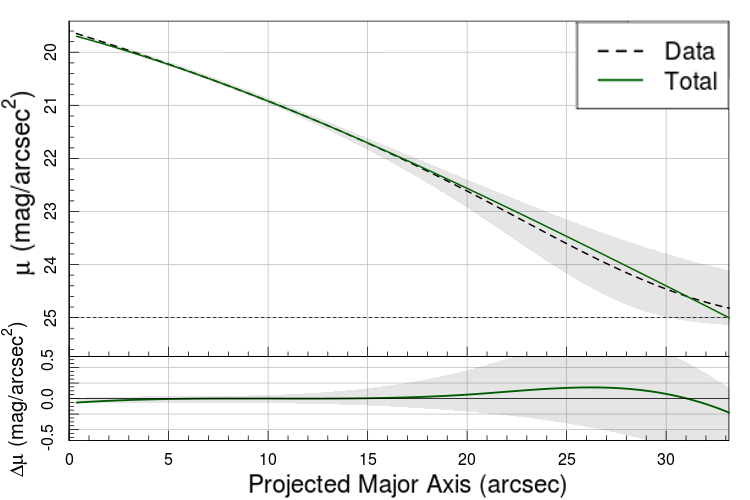}
    
    \includegraphics[width=0.625\textwidth]{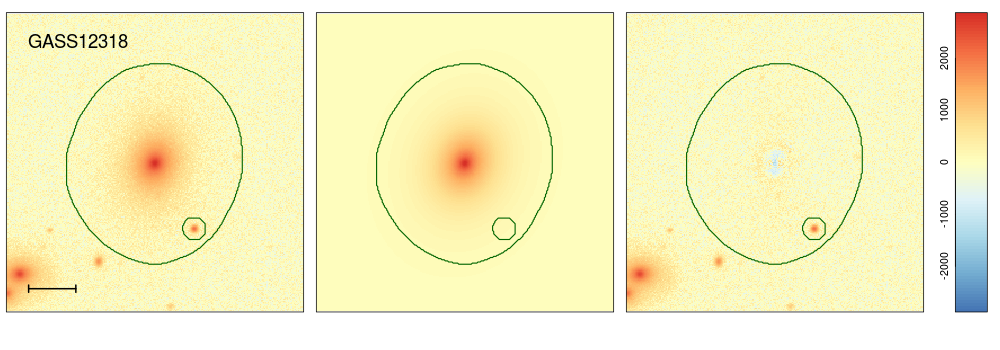}
    \includegraphics[width=0.325\textwidth]{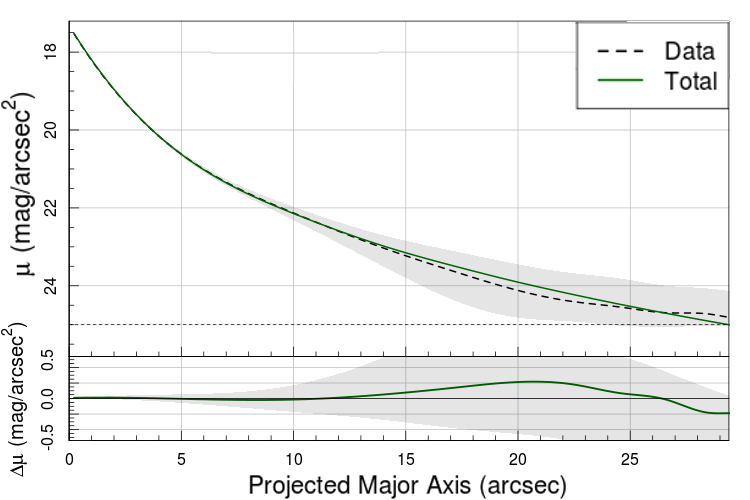}
    
    \includegraphics[width=0.625\textwidth]{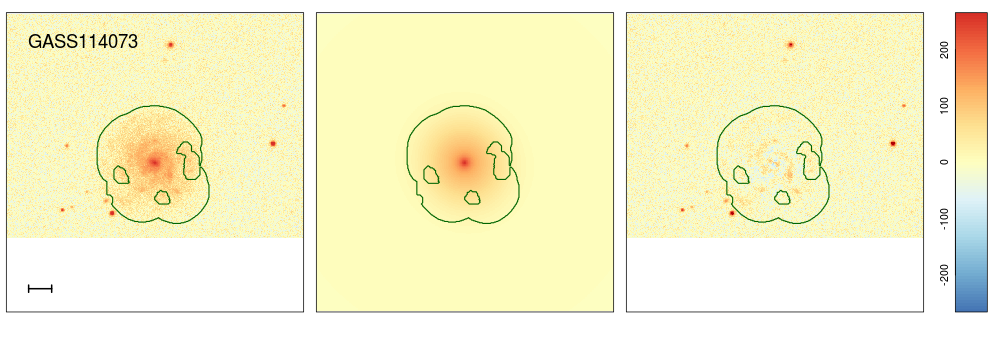}
    \includegraphics[width=0.325\textwidth]{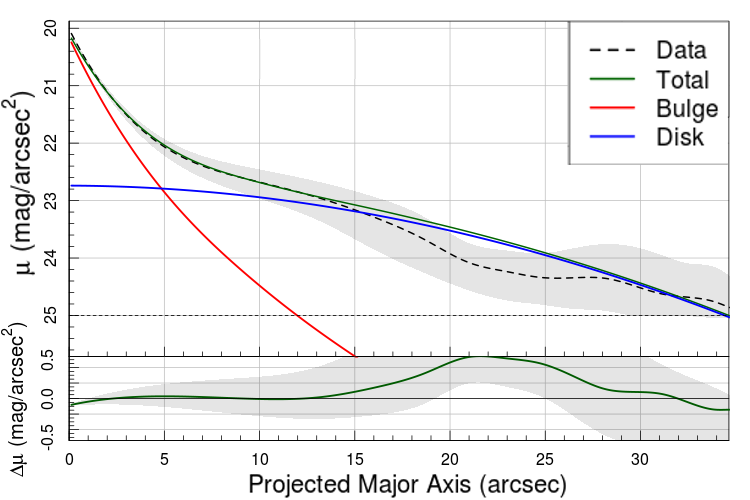}
    
    \includegraphics[width=0.625\textwidth]{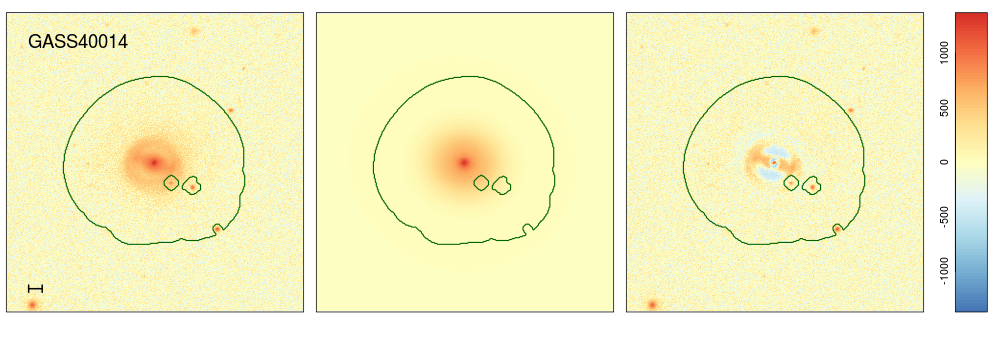}
    \includegraphics[width=0.325\textwidth]{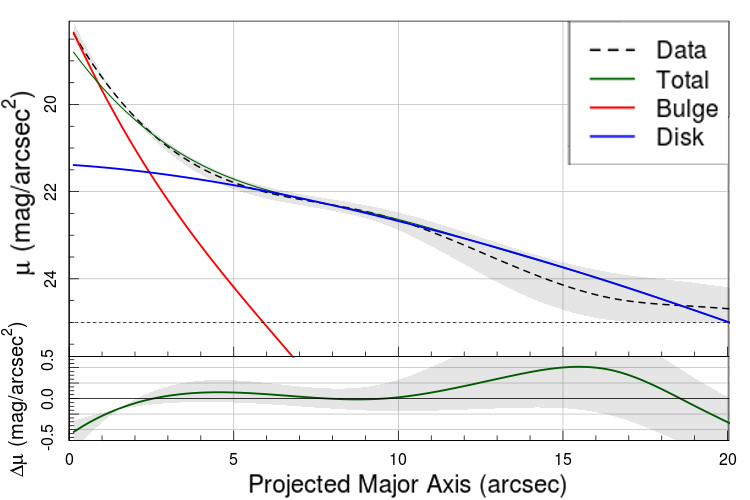}
    
    \includegraphics[width=0.625\textwidth]{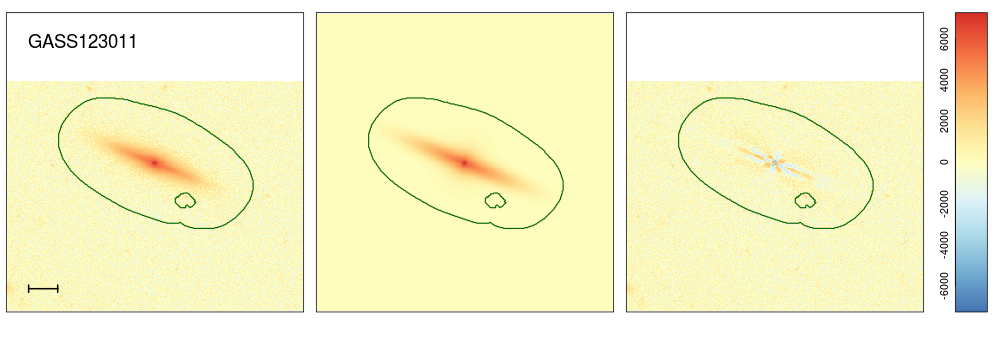}
    \includegraphics[width=0.325\textwidth]{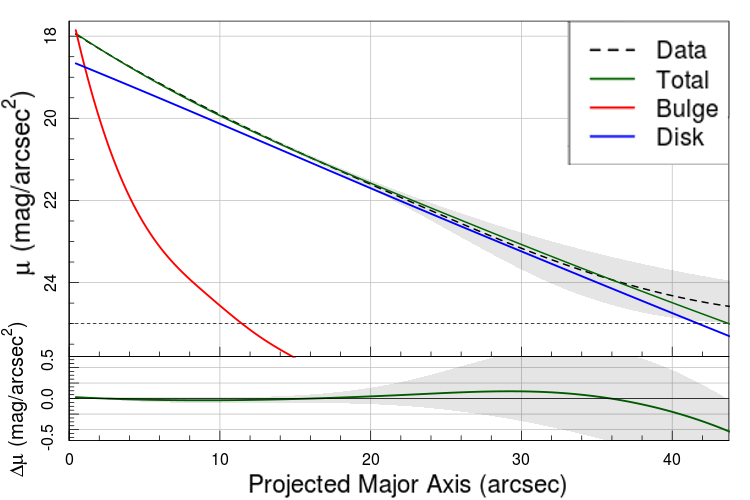}
    
    \caption{Examples of some fits obtained using \ProFit on \SDSS \textit{r}-band images; each row shows the result of a single galaxy. The panels show from left-to-right: the original \textit{r}-band \SDSS image centred upon each galaxy, the optimised model obtained from fitting either a single- or double-\Sersic model with \ProFit and the residual between both. The horizontal lines in the galaxy images denote a length of 5\,kpc at the distance of the galaxy. For the residual image, bluer pixels indicate regions that have are over-subtracted with respect to the image data and redder pixels indicate an under-subtraction. The azimuthally-averaged surface brightness profiles shown in the right panels are obtained by deprojecting the image pixels along the major axes of the bulge and disc components separately. For double-component models, red and blue lines show the bulge and disc component, respectively and the combined profile is shown in green; only the total profile is shown for single-component models. In all plots, the black dashed line shows a smooth spline representation of the image data and the accompanying shaded region shows the scatter about the deprojected profile. The lower panels of these plots show the one-dimensional residuals (data - model).}
    
    \label{fig:example_fits}
\end{figure*}

We run \ProFound on the \rband images to measure global parameters of the segment from which we can derive reasonable initial guesses for the model. In addition to the positional parameters (coordinates of maximum flux), we extract the axial ratio ($b/a$), positional angle ($\theta$) and size (as an effective radius, $R_{e}$) of a segment to inform upon the structural parameters. We then make use of the total flux and concentration index to make an informative estimate of the total magnitude and likely shape ($n$) of the light profile. Given these rough initial guesses, we fit a single \Sersic model to the data by first using a fast, but biased, gradient descent optimiser (specifically, a Levenberg-Marquardt algorithm). Provided the model parameters converge, this is then used as an initial guess for a more expensive, but significantly more robust, Markov Chain Monte Carlo (MCMC) optimiser which is used to accurately refine the parameter posteriors. We use a Component-wise Hit-And-Run Metropolis (CHARM) algorithm with $10^{4}$ iterations as this provides an exhaustive traversal of complex parameter spaces, good handling of highly correlated parameters and works reasonably well with multimodal distributions; all common features found in models of galaxy light profiles. When using $\chi^{2}$ minimisation routines, \citet{Lange2016} expressed the need for using a grid of initial guesses across the model parameter space in order to identify whether the fitting converges. By instead using an MCMC optimisation routine, the optimisation solutions are stable in that they are insensitive to the initial guesses chosen.

Once a galaxy has been fit with a single-component model, we adapt the optimised parameters (specifically, $n$, $R_{e}$, $b/a$) into a set of initial guesses for a more complicated bulge + disc model involving two superimposed \Sersic functions and run the optimisation stages as stated above. The disc component is modelled as a near-exponential with the freedom to explore \Sersic indices in the range of $0.5 \leqslant n_{Disc} \leqslant 1.5$. The intervals of the bulge \Sersic index begin in the range of $1.25 \leqslant n_{Bulge} \leqslant 20.0$ and, initially, the bulge component is forced to be concentric with the disc component. We also assume the bulge to be strictly spherical and hence we do not fit for its axial ratio ($b/a \equiv 0$) nor its positional angle. These constraints simplify the optimisation of the bulge parameters, further reducing the possibility of the bulge modelling a different region of the galaxy (e.g. underlying diffuse features, clumps or even the disc itself).

Finally, we \textit{bootstrap} the \textit{g}-band and \textit{i}-band models from the \textit{r}-band model by keeping all parameters of the double-component model fixed except for the positional parameters, the disc effective radius as well as the bulge and disc magnitudes. This naturally preserves the structure of the galaxy whilst allowing the colours of the bulge and disc to be accurately modelled. In the case of the single-component models, we allow all parameters to be free between bands to accommodate the relative contributions of the bulge or disc across the three \SDSS filters; in this case, the r-band model simply acts as a very reliable initial guess.

Figure \ref{fig:example_fits} shows some examples of the resulting models for several galaxies spanning a range of morphologies. Specifically, we show models for single-component fits (i.e. pure-disc and pure-spheroid) as well as double-component models for galaxies with distinct bulge and disc components in various configurations (face-on, barred spiral, edge-on). For the barred example in particular, the residual image shows that the residual flux is associated with the strong bar structure.

\subsection{Filtering and Model Selection}
\label{sec:model_filtering}
In this work, one of our goals is to study the cold gas properties of galaxies at fixed disc properties and so our profile fitting and model selection is optimised to recover the best parameters in the disc component. However, fitting models to the bulge and disc components of galaxies in such a way that is both automated and reliable has proven to be a notoriously difficult problem. Complications often emerge from artefacts in the data such as poor background subtraction, impinging unassociated sources, unreliable PSF models, etc. Furthermore, the parameter space of even a simple model can be multimodal, correlated and fraught with local minima. This often causes gradient descent optimisers to not necessarily converge or to explore non-physical regions of the parameter space (e.g. \Sersic indices of $n \sim 20$, effective radii as low as $R_{e} \lesssim 0.1''$). Even factors as fundamental as the distorted nature of some galaxies can impact the quality of the fit. One must consider effects from merging and tidally interacting galaxies, projection effects due to inclination especially where dust lanes are prominent, flocculant or clumpy light profiles, etc.

Given the above, it is inherently very difficult to define an automated process which decides the appropriate complexity (number of components) of the model, particularly given that the optimisation routine has no inbuilt knowledge of the characteristics of galaxies. Whilst a measure of a model's \textit{goodness of fit} (such as the reduced $\chi^{2}$, log-marginal likelihood) may be useful to determine whether the model is a \emph{statistically good} fit to the data, it does not express whether the model is a physically meaningful representation of a galaxy. Thus, whilst we do not utilise the model posteriors directly to select the best model, we retain them to help infer the quality of the fits and determine the errors in the model parameters.\\

Recently, studies such as \citet{Dimauro2018} have implemented machine learning algorithms to select the best fitting model based on photometric images. This is particularly useful when visual inspection is no longer feasible (see also, \citealt{DominguezSanchez2018, Tuccillo2018}). This technique, however, requires a large training set derived either from a much larger sample of visually inspected galaxies (e.g. Galaxy Zoo 2; \citealt{Willett2013}) or relying upon simulated synthetic galaxy images. These assumptions are not necessarily applicable for this sample of $\sim 1200$ galaxies at $z \sim 0$. Other studies, such as \citet{Allen2006}, introduced \emph{logical filters} to separate fits into particular classes based on the relative contributions of the bulge or disc at different radii as well as the structural information from the size and \Sersic index of the bulge component. This is indeed a good way to flag immediately spurious models: e.g. the bulge and disc components have swapped, the bulge component dominates the outer regions. However, it is difficult to decide what model is appropriate for galaxies that fall within these classes.

\begin{figure}
    \centering
    \includegraphics[width=\columnwidth]{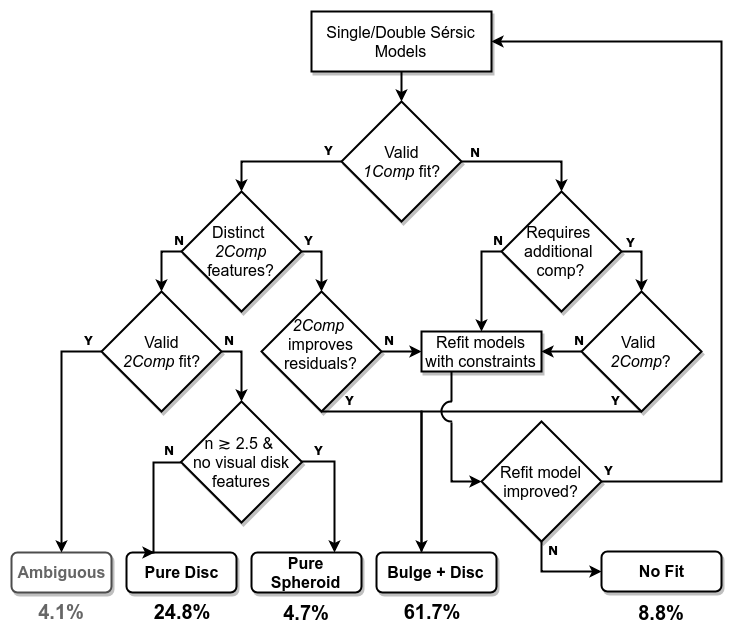}
    \caption{Decision tree used in the visually-guided inspection of galaxy models to validate and select the most appropriate model for each galaxy. A ``valid'' model is one which correctly models the light associated with the bulge and/or disc and the model parameters are physical. Galaxies are classified into pure-discs, pure-spheroids, composite bulge\,+\,disc models and those where no successful model could be obtained after constraints were added. The percentages indicate the proportion of galaxies that have been assigned to each class.}
    \label{fig:decision_tree}
\end{figure}

The relatively small size of the \xGASS sample (particularly in comparison with most all-sky surveys) warrants applying a similar logical filter based upon the optimised model parameters and, in addition, can be informed by the visual inspection of the residual data. The classification of galaxies required that the entire sample be visually confirmed according to the decision tree illustrated in Figure \ref{fig:decision_tree}; this was done independently between two authors (RC, LC) until agreement was met. This begins by checking whether the single-component model residuals and optimised parameters are a ``valid'' fit to the data in that they correctly model the light from the disc and/or bulge components. Secondary residual features that are not considered in our models including bars, rings, lenses, spiral arms, etc., should naturally remain in the residual images. If the single-component model is not valid, we check whether the missing component (often the bulge) is rectified in the double-component model, in which case the galaxy is classified as a bulge\,+\,disc system. If neither models are considered valid, we attempt to refit the galaxy with additional constraints that avoid the non-physical solution. If instead the single-component model is a valid fit to the data, we only consider this to be the best model for the data if the double-component model does not improve the residuals or there are no obvious visual indications of a distinct features associated with either a bulge or disc. In the presence of these features (e.g. spiral arms, bars, rings, etc.), if the double-component model fails to produce a physical solution, we do not revert to the single-component model but again attempt to refit the model with intuitive constraints (see below).

A galaxy identified as a single-component model must then be classified as either a pure-disc or pure-bulge system. This has traditionally been done using a cut in \Sersic index ranging between $n = 1.5$\,--\,$3.0$ \citep{Allen2006,Kelvin2012,Meert2015}. However, as the \Sersic index of a disc component is sensitive to disc breaks, central concentrations and emission in the outer regions, a low \Sersic index is not necessarily the definitive requirement of a pure-disc system. For this reason, we use $n \sim 2.5$ to roughly separate pure-discs from pure-bulges but inspect all galaxies to ensure that models with $n > 2.5$ but showing distinct disc features (spiral arms, low axial ratios, etc.) are correctly classified as pure-disc systems. In some cases, the choice between the single- and double-component models was considered ambiguous as both produced valid solutions and there were no discernible differences between their residual images. For these 48 (\simperc{4}) galaxies, all authors independently selected which model best fit the galaxy and the most preferred was chosen.

\begin{figure*}
    \centering
    \includegraphics[width=\textwidth]{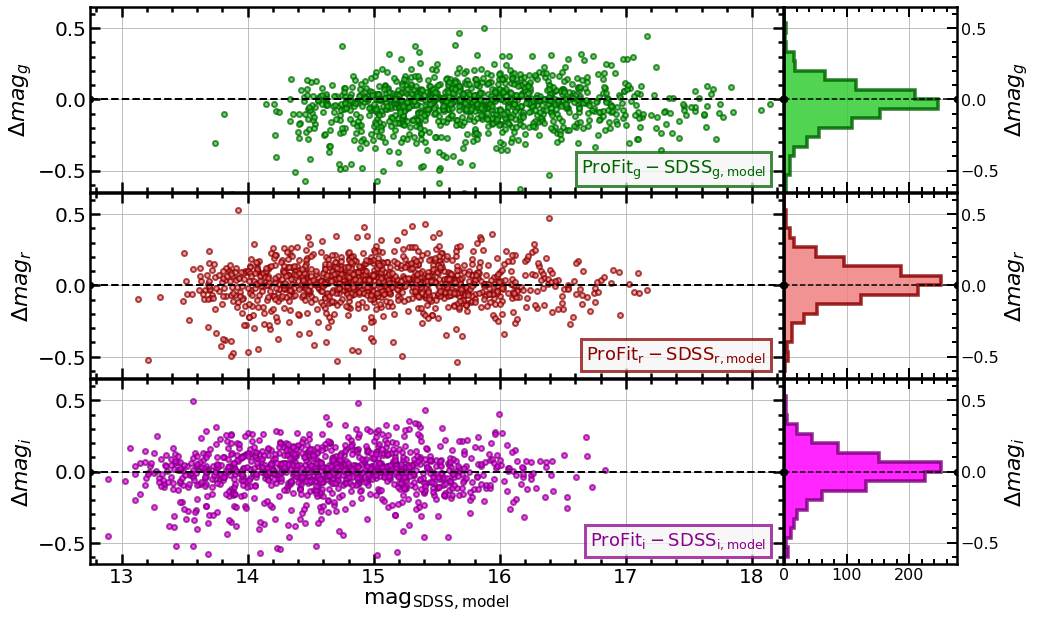}
    \caption{A comparison of the $g$, $r$ and \iband magnitudes measured in this study with those taken from the SDSS catalogue \citep{Stoughton2002}. The $\Delta mag$ quantity shows the difference in magnitudes of the \ProFit models from that of the \SDSS model magnitudes such that more negative values imply that our model recovers more flux. The histograms in the right panels show the spread in the differences of magnitudes.}
    \label{fig:mag_comparisons}
\end{figure*}

When modelling galaxies in \xGASS with \ProFit in an automated pipeline, we find that $\sim$80\% of the galaxies had converged upon a physically meaningful solution for at least one of the single- or double-component models. The remaining spurious fits need to be refit with added constraints to obtain a physical solution. There are many reasons for an unsuccessful fit to occur and most can be attributed to the fact that the symmetric single- or double-\Sersic profiles are more often than not an oversimplification of the complex geometries actually observed in galaxies. The potential shortcomings that were encountered from the automated fitting of models included:

\begin{enumerate}
    \item \textbf{An interaction with a nearby galaxy causing a distortion in the light profile.}\\
    Even if a pair of galaxies is not in a state of merging with one another, the overlapping light profiles can falsely imply some additional flux in the outer regions of the target galaxy; something that a high $n$, large $R_{e}$ \Sersic model (i.e. our bulge component) is capable of fitting. In such cases, it is often necessary to limit the upper boundary of the \Sersic index and impose that the bulge effective radius remains smaller than the disc.\\
    
    \item \textbf{Prominent secondary features which significantly influence either component.}\\
    Such secondary features can include: bars, spiral arms, lenses, rings, disc breaks, etc. It is indeed possible to include such features as additional components in a model, for instance, \citet{Laurikainen2004,Laurikainen2005,Gadotti2008,Gadotti2009,Salo2015} modelled bars using the modified Ferrer function, \citet{Kim2014,Gao+Ho2017} include features such as disc breaks and rings into their models using truncation functions, and even nonaxisymmetric features such as spiral arms can be modelled using Fourier mode transformations \citep{Rix+Zaritsky1995,Kim2017,Gao+Ho2017}. Whilst such models more accurately describe observed galaxy light profiles, the added complexity and computational cost make automation impractical in large galaxy samples.
    
    The alternative is to impose specific constraints on parameters, in particular, the PA, $b/a$, $R_{e}$ and $n$ to discourage the optimiser from exploring solutions that fit to these secondary features. An example of this can be seen in the fourth row of Figure \ref{fig:example_fits} where the axial ratio of the disc component had to be constrained to avoid fitting the prominent bar.
    
    \item \textbf{Offset between centres of components.}\\
    Whilst the peak centres of both the bulge and disc are most commonly concentric, distortions due to interactions or those seen in irregular galaxies can cause the two to be significantly offset. The first automated run forced the centres of the two components to be bound to one another so as to avoid the bulge fitting an undesired clumpy feature. We relax this constraint wherever the bulge and disc centres appear visually separated. Importantly, this regime is one reason why fitting galaxies in two dimensions must perform better than a projection of a galaxy along its major axis.\\
    
    \item \textbf{Bulge fitting unassociated light.}\\
    As we have allowed the bulge component to explore a wide range of \Sersic indices ($1.25 \leqslant n \leqslant 20.0$), it has more freedom to fit regions of the light profile that are not associated to the central bulge. For instance, light that exists in the outer regions of the segment either externally (nearby bright star or galaxy, residual sky background) or internally (diffuse spiral arms, tidal tails) can be incorrectly incorporated into the bulge model. To avoid this, we add a constraint that the bulge $R_{e}$ must be less than the disc $R_{e}$ and reduce the upper interval of the bulge \Sersic index until we cannot obtain a better solution with $n<6$. Enforcing an exponential disc was sometimes necessary to encourage the disc to incorporate the light that would otherwise be absorbed by the bulge.

\end{enumerate}

After rerunning the fitting process on these spurious fits with appropriate constraints, we result in a total of 1073 ($\sim 91$\,\%) galaxies that have been successfully modelled with at least a single- or double-component model. The remaining galaxies are those either too disturbed to model or unable to be improved with added constraints. We have identified 34 ($\sim 3$\,\%) galaxies which we deem not feasible to fit with our simplistic assumptions of a symmetric, monotonically-declining light profile. Nevertheless, the cold gas properties of such a disturbed system are likely very difficult to interpret.

\begin{figure*}
    \centering
    \includegraphics[width=\textwidth]{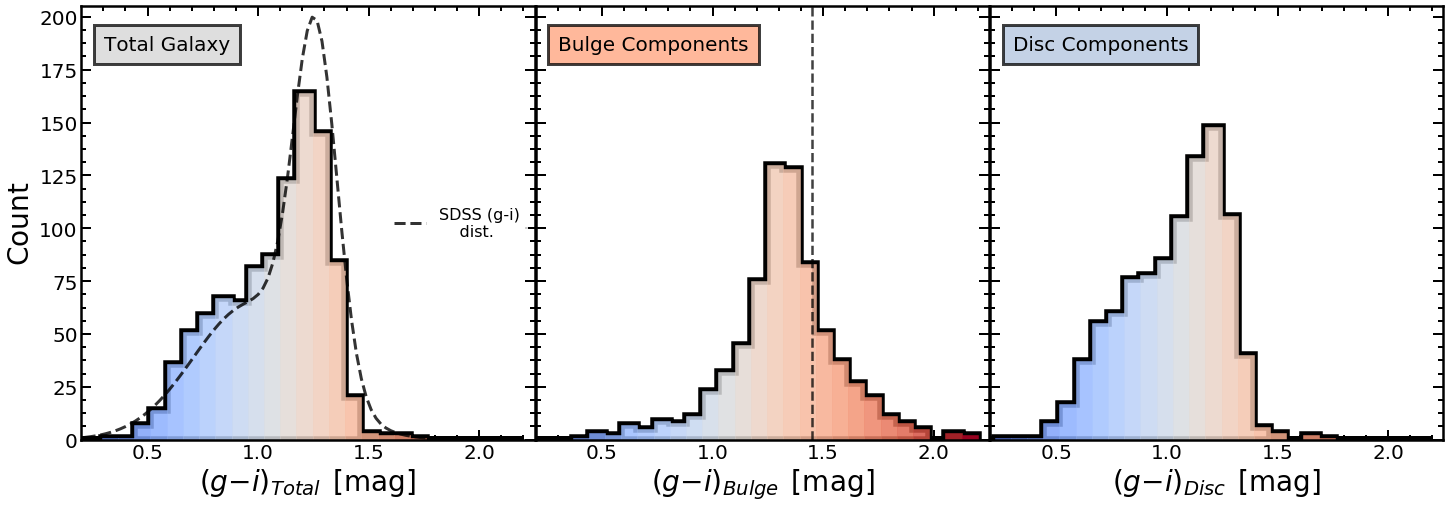}
    \caption{Distributions of the (\gi) colours for the summed modelled galaxy colour (left) as well as the bulge (middle) and disc (right) components taken separately. The distributions include both the single-component models (i.e. pure-bulges and pure-discs) taken as a whole as well as the bulge and disc components of the double-component models. The dashed curve in the left panel shows a bimodal Gaussian fit to the distribution of \SDSS (\gi) colours for the same sample of galaxies shown in the histogram. The vertical dashed line in the middle plot shows the colour limit imposed for measured bulge (\gi) colours in our sample.}
    \label{fig:component_colours}
\end{figure*}

\section{Validation of model parameters}
\label{sec:model_validation}
In this section we discuss the quality of the derived quantities from the structural decomposition of the \xGASS sample. We do this with reference to previous measurements of basic galaxy properties from the \SDSS catalogue as well as making direct comparisons with previous structural decompositions of this sample from \citet{Simard2011} and \citet{Meert2015}.

\subsection{Model Magnitudes}
\label{sec:magnitudes}
As a basis for the validation of our models, Figure \ref{fig:mag_comparisons} compares the total $g$-, $r$- and \iband magnitudes measured in our \ProFit models against the \SDSS model magnitudes. The \ProFit total magnitude represents either a single-component model magnitude or the summed magnitudes of the bulge and disc components for double-component systems. Overall, our $g$-, $r$- and \iband models exhibit average offsets from the \SDSS model magnitudes of -0.05, +0.01 and -0.01 mag and standard deviations of 0.27, 0.28 and 0.36 mag, respectively. This reinforces that there is to first order a good level of agreement in the photometric measurements between these two different methods. We see the greatest differences at the brightest end of the magnitude distribution where our models tend to recover slightly more flux. This is because the \SDSS model magnitudes are truncated beyond $7\,\times$ and $3\,\times$ the effective radius of the profile for de Vaucouleurs and pure-exponential profiles, respectively. Indeed, the galaxies with the largest offsets are indicative of higher \Sersic index models where the truncation will have the largest impact on the measured magnitude. Similar offsets and trends have also been noted in previous structural decomposition studies of both observations \citep{Hill2011,Kelvin2012,Meert2015} and simulated galaxy images \citep{Bernardi2014}.\\

\subsection{Separating Galaxy Components}
\label{sec:colours}
One of the main aims of this work is to understand the properties of galaxies through their individual components. In particular, the colours of bulges and discs give insight into the ages, metallicities and internal dust extinction of their stellar populations. Hence, measurements of component colours can be used to empirically derive estimates of the stellar mass to within a factor of $\lesssim 2$ uncertainty \citep{Zibetti2009}. Figure \ref{fig:component_colours} shows the distributions of total and individual component (bulge and disc) colours for the \xGASS models. Our models reproduce the nominal bimodality of galaxies into the blue and red sequences. As a reference, the dashed curve in the panel of total galaxy colours represents a double Gaussian fit to the distribution of \SDSS (\gi) colours for the same sample. It shows that the peaks of the distribution for the blue and red sequence are in good agreement. This follows from our relatively good agreement of the total magnitudes across all bands in Figure \ref{fig:mag_comparisons}.

The distribution of disc colours tends to span the bluer region in addition to a peak of discs that are slightly red. Conversely, the distribution of bulge (\gi) colours has a median value of $1.33$\,mag with extended tails towards both blue and red colours. The blue regime reflects a population of bulges more characteristic of pseudobulges; typical of flattened profiles (low $n$), generally more extended (large $R_{e}$) and exist predominantly in the lowest mass systems ($M_\mathrm{\star} \lesssim 10^{10}$\,\Msun). Their SDSS composite images confirm that these bulges are indeed blue rather than an artefact of our model fitting. The tail of bulge colours out to excessively red (\gi) colours has a less clear explanation; values as high as $(g-i) \gtrsim 1.5$\,mag do not appear physically plausible \citep{FernandezLorenzo2014,Mendel2014,Kim2016}. These bulges are still likely to be some of the most red objects in the sample but our ability to derive a meaningful quantity is severely lessened in the regime where the bulge magnitudes in the $g$-band are significantly fainter than in the $i$-band. As our profile fitting is optimised to obtain the best parameters for the disc, the modelled parameters of bulges are expected to be far less robust. In addition, a large fraction of the bulges (\simperc{25} of double-component galaxies) are not sufficiently resolved ($\lesssim\,$40\% of the PSF's FWHM) to obtain meaningful structural parameters; e.g. $R_{e,bulge}$, $n_{bulge}$, $b/a_{bulge}$; \citep{Gadotti2008,Meert2013,Bernardi2014}. As we discuss later, this has important implications for quantities involving the bulge stellar mass such as the bulge-to-total stellar mass ratio. Discs, however, are well resolved in this sample and thus their model parameters are more reliable. Indeed, as expected, we recover a relatively flat distribution in measured disc axial ratios.

\subsection{Comparisons with Previous Literature}
\label{sec:literature_comparisons}
The bulge-to-total ratio is an important quantity in defining the internal structure or morphology of a galaxy in a way that is not dependent on a visual classification scheme. However, accurately modelling the bulge fraction in terms of a particular optical filter ($\beta_{x}$) or, alternatively, in terms of the component stellar masses (\BtoTM) has proven to be a difficult challenge in many structural decomposition studies. There are many sources of uncertainty involved in both the fitting of light profiles as well as in the model filtering and selection. In this section, we compare our results to those of \citet{Simard2011} and \citet{Meert2015} as, despite there being numerous methodological differences, they draw the most parallels with this work. In particular, these are based on SDSS photometric images and fit double-component free $n_{Bulge}$ \Sersic with an exponential disc (albeit we allow some freedom such that $0.5 < n_{Disc} < 1.5$).

\begin{figure}
    \centering
    \includegraphics[width=\columnwidth]{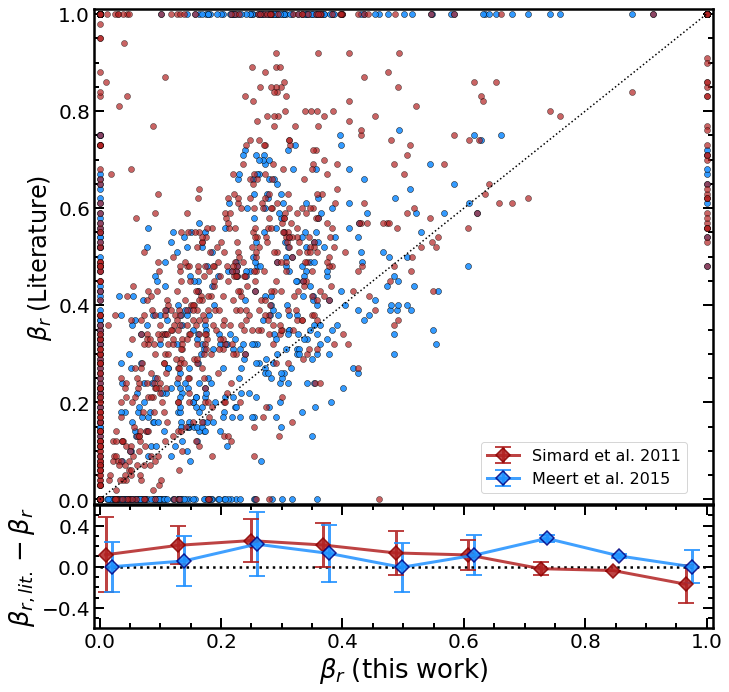}
    \caption{Comparison between the $r$-band bulge-to-total flux ratios (\BtoTr) modelled in \citet{Simard2011} (red) and \citet{Meert2015} (blue) against those measured in this work. We compare only galaxies that are considered \emph{valid} in the respective catalogues (see text for details). The bottom panel shows the median difference between the literature measurements and this work as a function of our \BtoTr measurements. The error bars show the 1$\sigma$ standard deviation of the differences.}
    \label{fig:literature_B2T_comparisons}
\end{figure}

For comparing with \citet{Simard2011}, we show the $r$-band bulge-to-total flux ratio (\BtoTr) in Figure \ref{fig:literature_B2T_comparisons}. This comparison includes galaxies that are considered either a single-component (i.e. pure-disc or pure-spheroid) or double-component (i.e. bulge + disc) system in \citet{Simard2011}. Double-component models are defined wherever the resulting \emph{F}-test value from comparing the $\chi^{2}$ residuals of their two models meets the condition that $P_{pS} \leqslant 0.32$. This is the threshold probability at which they define a pure-\Sersic model ($P_{pS}$) as being preferred over a more complex \Sersic\,+\,exponential model. The remaining pure-\Sersic models are classified as pure-discs ($\beta_{r} = 0$) if their \Sersic index is $n_{pS} < 2.5$ and a pure-spheroid ($\beta_{r} = 1$) otherwise. This ad hoc cut in $n_{pS}$ follows from previous literature studies but the comparison does not change dramatically if the cut is varied slightly.

We also compare to the \BtoTr values derived in \citet{Meert2015} by choosing double-component systems where their \textit{finalflag} of 10 is set (\textit{good} double-component galaxies) and single-component systems if the \textit{finalflag} has 1 or 4 set for pure-discs and pure-bulges, respectively (see \citealt{Meert2015} for more details on fitting flags). Removing galaxies identified as having problematic fits in \citet{Meert2015} leaves 894 galaxies which may be compared with the models obtained in this study.

Figure \ref{fig:literature_B2T_comparisons} shows that both \citet{Simard2011} and \citet{Meert2015} tend to measure systematically higher bulge-to-total ratios than in this work. The panel below suggests that we are, in general, in better agreement with \citet{Meert2015}, albeit, both still exhibit a relatively large scatter. The commonality is highlighted not only by a smaller offset in the average difference but, more importantly, by a similar distribution of this parameter space. \citet{Meert2015} also see a lack of galaxies with a very-high bulge-to-total ratios ($\beta_{r} \gtrsim 0.7$), further suggesting that this is a poorly constrained region of the parameter space. Notwithstanding the systematically lower bulge-to-total ratios measured in our sample, our greatest tension stems from the region of galaxies which we consider disc-dominated but have \BtoTr $\gtrsim 0.75$ (implying substantially bulge-dominated systems) in the previous literature. Interestingly, we note that this population has a generally blue total (\gi) colour and their composite colour images show that they are more indicative of disc-dominated systems.

The discrepancy is most evident in the \citet{Simard2011} models where it has been shown in previous studies \citep{Bernardi2014,Meert2015} that the systematic errors in the behaviour of the \Sersic index of the bulge models is due to over-subtraction of the background sky level. Furthermore, they have incorporated a strong prior on the \Sersic index of their bulge-components that effectively favours a de Vaucouleurs, $n = 4$ profile. This makes the comparison with \citet{Simard2011} less meaningful. \citet{Meert2015} do not impose this prior on the \Sersic index and as a result, we observe a very similar distribution in the \Sersic index of the bulge. These large model-based discrepancies reflect the necessity of an additional visual inspection stage capable of filtering spurious fits and allowing for remodelling with guided constraints. Such a process is, of course, not feasible in samples as large as those in \citet{Simard2011} and \citet{Meert2015} without employing numerous observers similar to that of the Galaxy Zoo project \citep{Lintott2008} or implementing deep learning algorithms \citep{Dimauro2018,Tuccillo2018}.

Despite the dramatic offsets caused by the particulars in choosing the most appropriate model, we still see a large scatter in the measured bulge-to-total ratio for the population of galaxies that are identified as being best modelled by a bulge\,+\,disc model. Some of this scatter is due to the different optimisation algorithms used in these studies that are subject to finding solutions in local minima, thereby increasing the measurement uncertainties \citep{Haussler2007,Gadotti2009,Meert2013}. This scatter is also driven by the different treatment of modelling discs to accommodate the non-exponential features of observed light profiles \citep{Pohlen+Trujillo2006}. As expected, galaxies modelled with $n_{disc} < 1$ tend to have higher bulge-to-total ratios as less light is attributed to the disc and conversely for galaxies modelled with $n_{disc} > 1$. Using the structural parameters measured in these studies does not qualitatively change the conclusions of our analysis. However, the scatter significantly increases mainly due to the presence of spurious fits, e.g., galaxies erroneously classified as bulge-dominated at low stellar masses.

\subsection{Decomposition of Galaxy Stellar Masses}
\label{sec:stellar_masses}
With the magnitudes of the bulge and disc components modelled separately in three of the SDSS bands, we compute their stellar masses separately and derive the total galaxy mass from their sum as well as directly from the total modelled magnitudes. We compute effective stellar mass-to-light ratios using the relations for optical colours given in \citet{Zibetti2009} which are derived through stellar population synthesis modelling. The equation is given by:

\begin{equation}
    \mathrm{log}\,M_{\star}/L_{r} = -0.977 + 1.157\,(g-i)
    \label{eqn:mass_to_light}
\end{equation}

\noindent
We adopt the \rband luminosity as this is the band from which our models are calibrated with additional filtering and validation. We choose the (\gi) colour as it provides a reasonably wide leverage across wavelength making it more sensitive to differences in the stellar populations of bulges and discs. This stellar mass estimate assumes that the initial mass function is universal and is well described by \citet{Chabrier2003}. We have experimented with various recipes for estimating the stellar mass-to-light ratios including different photometric bands and colours as well as a different prescription defined in \citet{Taylor2011}. The differences between the various stellar mass estimates are always less than $0.1$\,dex in the measured stellar mass and so the particular choice does not change our conclusions.

\begin{figure}
    \centering
    \includegraphics[width=\columnwidth]{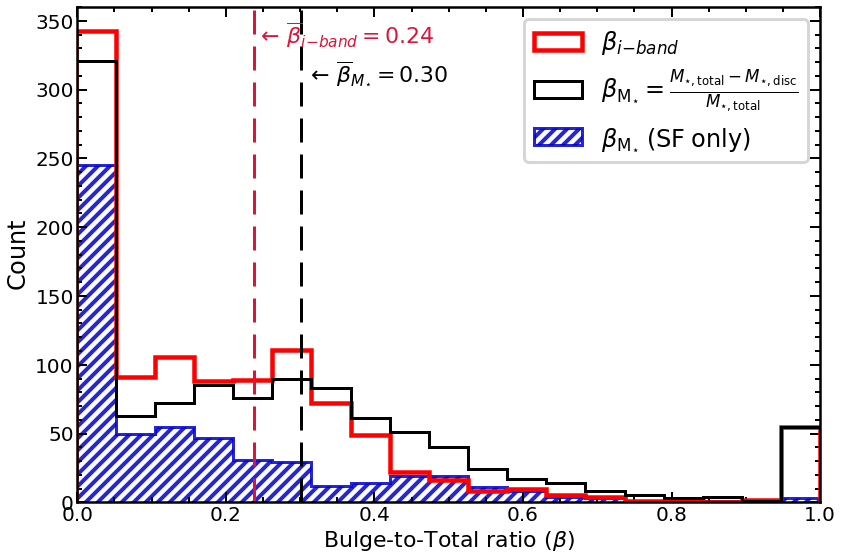}
    \caption{The distributions of both $i$-band (red) and stellar mass (black) bulge-to-total ratios measured in the \xGASS sample. The stellar mass bulge-to-total ratio is derived according to equation \ref{eqn:B2T_M}. The blue hatched histogram represents the distribution of stellar mass bulge-to-total ratios for the subset defined as star-forming. The dashed vertical lines represent the mean \BtoTi and \BtoTM of the double-component population.}
    \label{fig:B2T_distributions}
\end{figure}

We derive the total stellar mass of a galaxy based upon the global \rband magnitude and (\gi) colour measured from summing the bulge and disc magnitudes. The alternative to this involves summing together the stellar masses derived from the individual component colours. The two methods are quantitatively comparable with a negligible offset and a scatter of 0.03\,dex in their difference. However, the greater uncertainties in determining the (\gi) colours for the components individually (particularly the bulge) makes the summed stellar masses less reliable.\\

\begin{figure*}
    \centering
    \includegraphics[width=\textwidth]{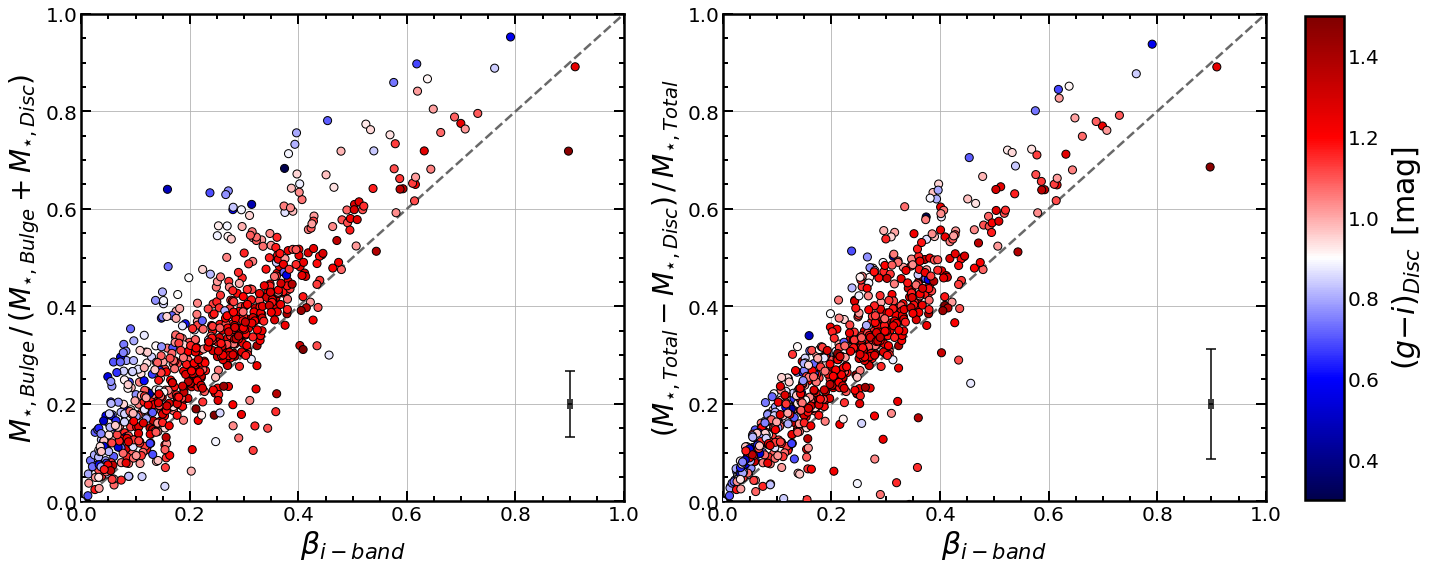}
    \caption{A comparison between the photometric \iband bulge-to-total ratio (\BtoTi) with both the stellar mass bulge-to-total taken from the component masses (left panel) as well as using the proxy bulge mass and the total galaxy mass (right panel). The points are coloured according to the measured (\gi) colour of the disc. The error bars in the bottom right indicate the average errors of the measured bulge-to-total ratio quantities.}
    \label{fig:B2T_mass_and_light}
\end{figure*}

Whilst the photometric bulge-to-total ratio provides a relative measure of how disc-dominated or bulge-dominated a galaxy appears, the \emph{stellar mass} bulge-to-total should be a more useful physical quantity for understanding the impact of the bulge. Figure \ref{fig:B2T_distributions} shows the distributions of both the $i$-band (\BtoTi) and stellar mass (\BtoTM) bulge-to-total ratios in the \xGASS sample. Here, we define the stellar mass bulge-to-total ratio as:

\begin{equation}
    \beta_\mathrm{M_{\star}} \equiv \frac{M_\mathrm{\star,total} - M_\mathrm{\star,disc}}{M_\mathrm{\star,total}}
    \label{eqn:B2T_M}
\end{equation}

\noindent
We assign the residual mass from subtracting the disc component away from the total galaxy mass to the bulge (i.e. $M_\mathrm{\star,bulge} \equiv M_\mathrm{\star,total} - M_\mathrm{\star,disc}$). In this way, we still utilise the information derived from the bulge and disc models but in a way that is less heavily biased by their measured (\gi) colours (see below).

The two peaks in Figure \ref{fig:B2T_distributions} at $\beta = 0$ and $\beta = 1$ are the population of single-component models corresponding to pure-discs and pure-bulges, respectively. The population of double-component models exist in between with averages of \BtoTi = 0.24 and \BtoTM = 0.30. In general, we see that the bulge-to-total ratio measured in terms of a galaxy's \emph{stellar mass} is larger than when measured with \emph{photometry} alone. This is due to the fact that bulges typically contain an older, more-massive population of stars whereas the disc predominantly contains the younger, less-massive stars. We see a tail in the distribution out to $\beta \sim 0.75$ with very few galaxies beyond this. This reflects two important points about both the sample and the methodology. Firstly, systems with bulges that encompass $\gtrsim70$\% of a galaxy's mass with a small, embedded disc structure are likely very rare objects indeed and thus --- in the small cosmological volume of the \xGASS sample --- we would not expect them to be numerous. Secondly, of these systems that are present, our ability to detect the diffuse, embedded disc would be severely limited by both the surface brightness limit of our data and how well the optimisation routine could recover sensible parameters. As such, we observe a void of galaxies in the high $\beta$ region as they move into the single-component pure-bulge population. The same is also true for the very faint bulges in the low $\beta$ regime, albeit to a lesser extent given that bulges are generally more centrally concentrated and easier to distinguish. Note that plotting the bulge-to-total ratio on a linear scale can be misleading as the dynamic range of this quantity is logarithmic. We recognise the discontinuous nature of this parameter space as one of the limitations in measuring the relative structure of galaxies. Figure \ref{fig:B2T_distributions} also shows the distribution of stellar mass bulge-to-total ratios for the subset of star-forming galaxies which we investigate further in Section \ref{sec:role_of_bulges_in_SFGs}. As expected, the distribution of \BtoTM for star-forming galaxies is skewed towards disc-dominated systems.

The method of computing \BtoTM is in contrast to using the individual component masses as is commonly done for the luminosity-based ratio. Figure \ref{fig:B2T_mass_and_light} shows a direct comparison between \BtoTM and \BtoTi where the left panel corresponds to the ratio of the individual component masses and the right panel to the residual bulge mass. Galaxies are coloured by the (\gi) colour measured from their disc models in order to show this bias. It follows from equation \ref{eqn:mass_to_light} that galaxies with the greatest difference between their bulge and disc colours will see the largest difference from \BtoTi to \BtoTM. Using the component stellar masses alone (left panel), the population that extends to higher \BtoTM contains not only galaxies with a significantly bright bulge, but also galaxies with very blue discs. This has the effect of lowering the amount of mass attributed to the disc and in turn drives up the bulge-to-total ratio. On the other hand, this effect is not as prominent in the right panel, where the bulge mass is taken as the difference between the total and disc stellar masses. In both methods, the galaxies that fall below the one-to-one line are typically those with a relatively blue bulge component and have structural features also similar to that of a pseudobulge. It is worth noting that the uncertainties in the stellar mass bulge-to-total ratios are often more than an order of magnitude greater than the luminosity bulge-to-total ratio. The average uncertainties are $0.07$ for the component \BtoTM (left panel) and $0.11$ for \BtoTM calculated from the residual bulge mass (right panel) whereas the average uncertainty in \BtoTi is 0.004.

This plot shows that the way in which we choose to define the bulge fraction in a galaxy will influence our interpretation of how the presence of a bulge affects the gas properties of galaxies in the following sections. For instance, when measuring \BtoTM from the individual component masses, the higher \BtoTM regime becomes contaminated by very star-forming discs and hence will likely also be quite \HI-rich. This quantity seems to say less about the presence of the bulge and more about the current star formation in the disc.

\begin{figure*}
    \centering
    \includegraphics[width=\textwidth]{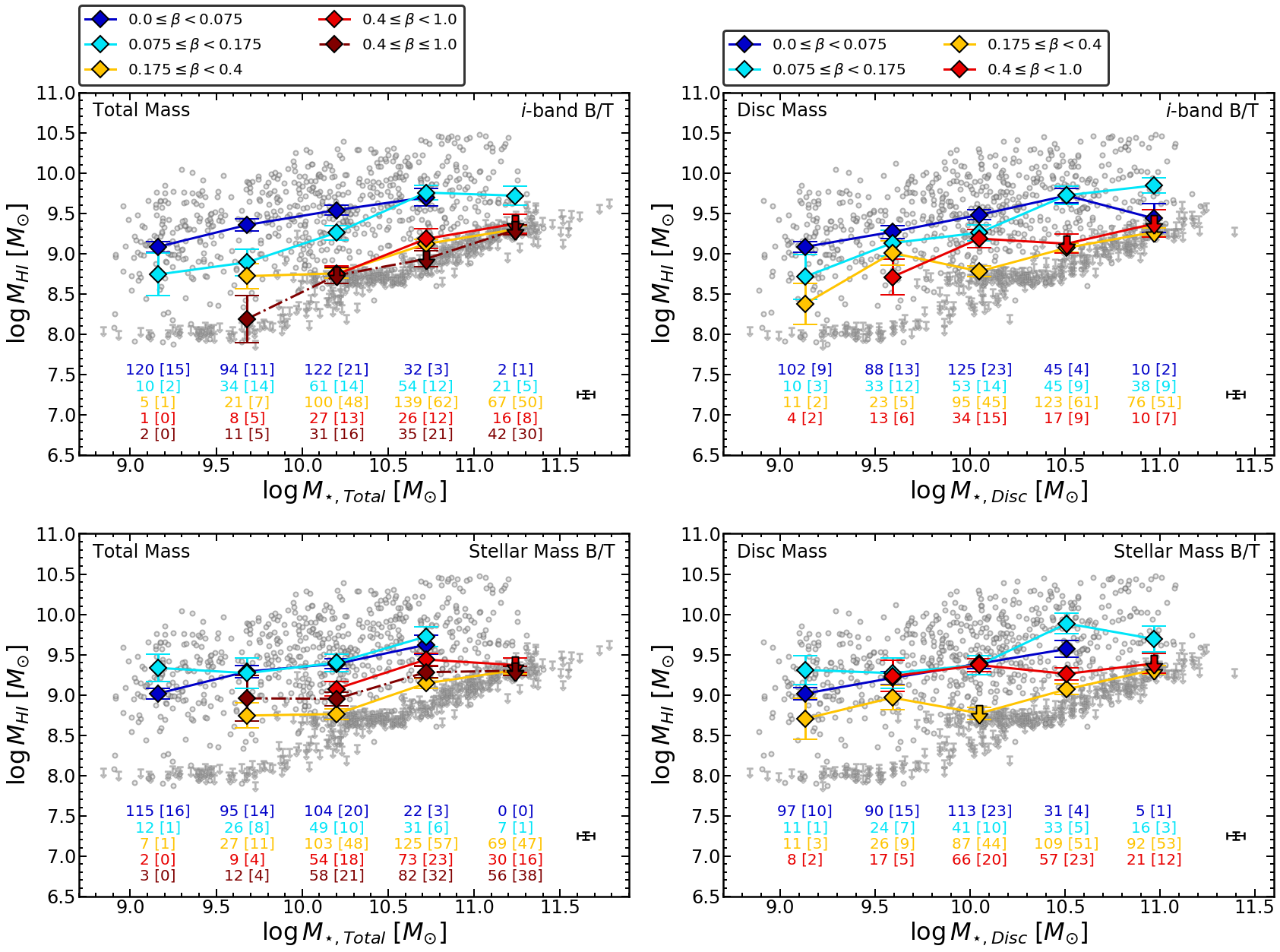}
    
    \caption{The \HI gas mass (\MHI) against the total (left) and disc (right) stellar masses separated into bins of i-band bulge-to-total ratio (\BtoTi; top) and stellar mass bulge-to-total ratio (\BtoTM; bottom). The different bins of bulge-to-total ratio were chosen to accommodate the logarithmic distribution of this parameter. The \emph{dark} red bin differs from the \emph{light} red bin by also including pure-bulges ($\beta = 1$); this is not possible in the panels plotted against disc stellar mass. Smaller circles and downwards arrows show \HI detections and non-detections, respectively. Error bars show the standard error on the median within each bin. Bins that are dominated by \HI non-detections are considered median upper limits and indicated as a large downwards arrow. The numbers below indicate both the total number of points in each bin and (in brackets) the number of those that are \HI non-detections. Only bins with at least ten galaxies contributing are plotted. The typical measurement errors in stellar mass and \HI mass (detections only) for individual points is shown in the bottom-right. The medians shown here are also provided in Tables \ref{tab:MHI_vs_Mstar_B2T_i} and \ref{tab:MHI_vs_Mstar_B2T_M}.}
    \label{fig:MHI_vs_Mstar_by_B2T}
\end{figure*}

\section{Results}
\label{sec:results}
We have shown in the previous sections that using \ProFit in combination with guided filtering and constrained remodelling, we have produced a robust catalogue of structural parameters for galaxies in the \xGASS sample. We now focus on understanding whether the presence of a galactic bulge affects the \HI reservoirs in galaxies.

\begin{figure*}
    \centering
    \includegraphics[width=\textwidth]{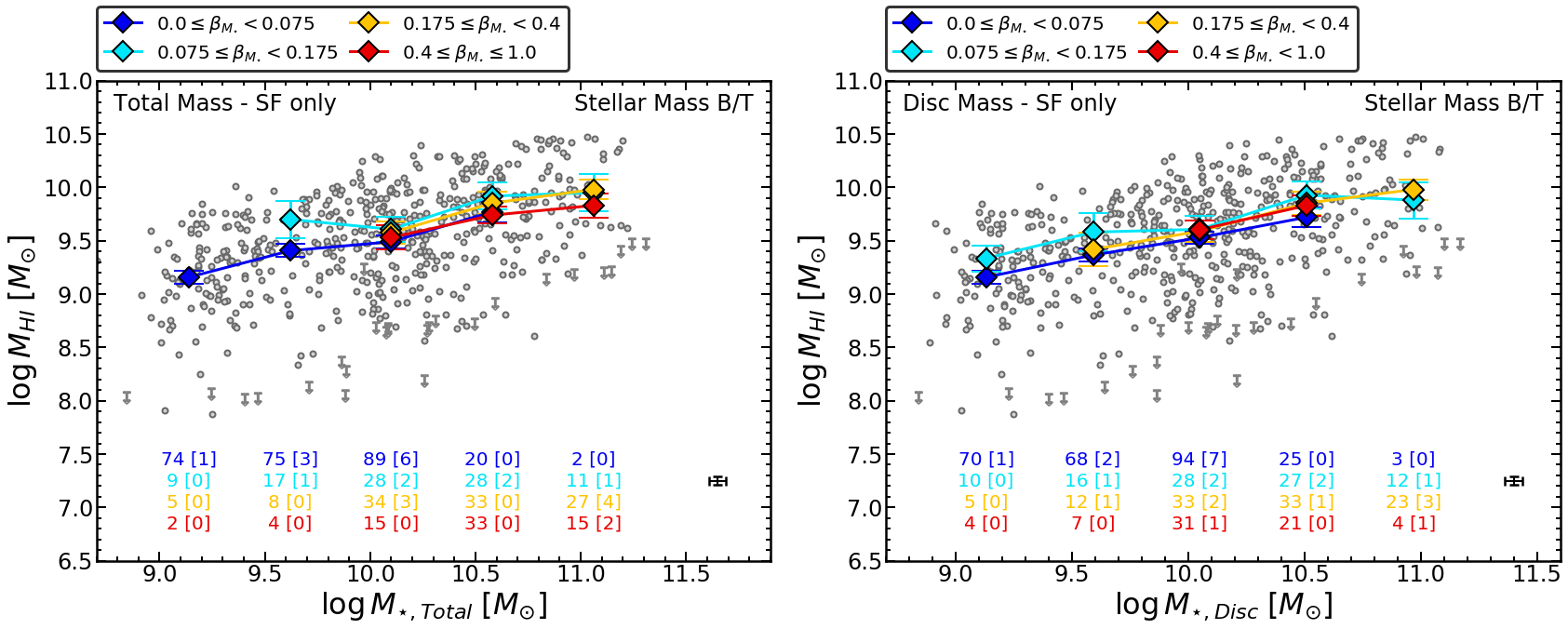}
    \caption{The same as Figure \ref{fig:MHI_vs_Mstar_by_B2T} but for the subset of ``star-forming'' galaxies based on them having a SFR that is higher than $1.5\times\sigma$ below the main sequence. Only the bulge-to-total ratio in terms of the stellar mass (\BtoTM) is shown here. The medians shown here are also provided in Table \ref{tab:MHI_vs_Mstar_B2T_M_SF}.}
    \label{fig:MHI_vs_Mstar_by_B2T_SFonly}
\end{figure*}

\subsection{The Impact of Bulges on Gas Scaling Relations}
\label{sec:bulges_in_gas_reservoirs}
In this section, we investigate the common scaling relation between atomic gas mass and stellar mass \citep{Catinella2018} for galaxies with different bulge-to-total ratios. As it is reasonable to assume that the cold gas properties of galaxies are associated with their disc components, we also compare the \HI mass at a fixed disc mass (\MstarD). Figure \ref{fig:MHI_vs_Mstar_by_B2T} shows a series of \MHIvM relations for both \MstarT and \MstarD separated into bins of bulge-to-total ratio. We first investigate the presence of any secondary trends with the photometric $i$-band bulge-to-total ratio in the top panels. As the distribution of the bulge-to-total ratio covers a logarithmic dynamic range (see e.g. Figure \ref{fig:B2T_distributions}), the $\beta$ bins are divided accordingly into logarithmically sized steps. Note also that the \emph{light red} and \emph{dark red} bins differ only in that the latter also includes pure-bulge systems which cannot be represented in the panels plotted against disc mass. The larger diamond points indicate the median \HI masses across stellar mass for the different bins in bulge-to-total ratio. As these are medians, as long as the median value is a detection, they are not affected by the exact treatment of the upper limits for galaxies not detected in \HI. If the median is itself a non-detection, it should be considered as an upper limit, as indicated by a large downwards arrow in Figure \ref{fig:MHI_vs_Mstar_by_B2T}. In general, we see that the lowest \BtoTi bin ($\sim$\,pure-discs) lie systematically above the other bins whereas the highest \BtoTi bin (bulge-dominated) consists mostly of \HI non-detections and hence follows closely with the gas fraction limit(s) of \xGASS. This first panel shows a fairly clear set of parallel relations for increasing bulge fractions. Moreover, we also see a shift in the locus of each \BtoTi bin across stellar mass showing that disc-dominated galaxies are seen mostly at low masses and bulge-dominated galaxies at high masses. This division is not seen as clearly in previous structural decomposition studies.

In moving to the panel plotted instead against the disc mass (top right), the positions of galaxies in the lower two \BtoTi bins will, by definition, change very little as the vast majority of their total mass is already in the disc. The higher \BtoTi bins on the other hand will shift (leftward) the most, causing the division between the \BtoTi bins to become slightly narrower. Note that the highest \BtoTi bin can no longer include pure-bulge galaxies (dark red dashed lines in \MstarT plot) as by definition they have no mass in a disc component ($\beta \equiv 1.0$). That said, removing the $\beta \equiv 1.0$ galaxies from the \MHI vs. \MstarT plot does not entirely explain the differences in the separations between these plots. In both plots we see a trend for galaxies with the lowest bulge-to-total flux ratios to be the most gas-rich across the entire range of stellar mass. The fact that this trend becomes less distinct when observing the relation at a fixed disc mass may suggest that the discs are more alike in terms of their gas properties than the galaxy treated as a whole.

The bottom two panels of Figure \ref{fig:MHI_vs_Mstar_by_B2T} show respectively the same quantities on both axes as the panels above but are instead binned by the stellar mass bulge-to-total (\BtoTM) given by Equation \ref{eqn:B2T_M}. Note that the distributions of both bulge-to-total ratio prescriptions cover slightly different ranges (see Figure \ref{fig:B2T_distributions}). Here, by using \BtoTM to designate the relative prominence of a bulge, we see a slight difference in the trends of bulge-fraction with \MHI across \Mstar. The offset for disc-dominated systems to have higher gas masses is less obvious when split by \BtoTM; in fact, the two lowest \BtoTM bins share nearly the same relation. This tells us that the bulge-fraction as measured as a ratio of fluxes or of stellar masses does not have the same physical interpretation; galaxies with a high \BtoTM are not necessarily equivalent to galaxies of similarly high \BtoTi. It is clear that the two definitions cannot be used interchangeably. When observing this at a fixed disc stellar mass, the separation between bulge-to-total bins becomes very small at low stellar masses. In fact, it seems that below $\sim 10^{10}$\,\Msun, there is a trend for the most bulge-dominated galaxies to lie systematically above the intermediate bin ($0.175 \leqslant \beta_\mathrm{M_{\star}} < 0.4$). The relation of the highest \BtoTM bin matches closely with that of the disc-dominated bins.\\

\subsection{Structure or Star Formation?}
\label{sec:role_of_bulges_in_SFGs}
Embedded within the scatter of the panels in Figure \ref{fig:MHI_vs_Mstar_by_B2T} is the residual trend of star formation activity with morphology \citep{Williams2010}. As has been highlighted by many previous studies \citep{Franx2008,Kauffmann2012,Cheung2012}, disc-dominated galaxies tend to be more actively forming stars whereas the bulge-dominated population is typically more passive. It is thus important to control for this observed correlation in order to understand whether the presence of the bulge is directly influencing the cold gas reservoirs or the trends we see are instead a consequence of properties that govern the star formation rate. The \xGASS sample is not large enough to define statistically significant bins simultaneously in both \SFR and $\beta$. Hence, in Figure \ref{fig:MHI_vs_Mstar_by_B2T_SFonly}, we instead focus only on the population of star-forming galaxies. This excludes galaxies that have already undergone a quenching process; isolating only galaxies where the process of converting their cold gas reservoirs into new stars is still ongoing. In doing so, we also mitigate the impact, if any, in how we treat the \HI non-detections in our sample as there are few examples of quenched, \HI-rich galaxies. Figure \ref{fig:B2T_distributions} shows also the distribution in \BtoTM for this subset of star-forming galaxies showing, as expected, that they are preferentially disc-dominated. We define the subset of star-forming galaxies in \xGASS according to their \emph{distance} away from the star-forming main sequence (SFMS) as defined in \citet{Catinella2018}; see also \citet{Janowiecki2019}. This yields the following expression for the \xGASS sample:

\begin{equation}
    {\rm log}\,(sSFR_\mathrm{MS}) = -0.344\left({\rm log}\,(M_{\star}) - 9\right) - 9.822
    \label{eqn:SFMS}
\end{equation}

\noindent
with a corresponding standard deviation as a function of stellar mass given by:
\begin{equation}
    \sigma_\mathrm{MS} = 0.088\left({\rm log}\,(M_{\star}) - 9\right) + 0.188
    \label{eqn:sigmaSFMS}
\end{equation}

\noindent
The panels in this figure show the same quantities as the lower panels of Figure \ref{fig:MHI_vs_Mstar_by_B2T} but only for the subset of galaxies with $sSFR \geq sSFR_{MS} - 1.5\times\sigma_{MS}$. This shows that the separation of median \MHI at fixed \Mstar between different bulge-to-total ratios is negligible when considering star-forming galaxies only. The error bars represent the standard error on the medians. This verifies that the small differences are within the errors on the measurements. Conversely, using instead the complement of this star-forming subset (i.e. $sSFR < sSFR_{MS} - 1.5\times\sigma_{MS}$) also shows no trend with the bulge-to-total ratio, albeit this regime is mostly dominated by \HI non-detections. This result holds for different cuts from the star-forming main-sequence as well as when selecting galaxies on their \NUVr colour (not shown).

\begin{figure}
    \centering
    \includegraphics[width=\columnwidth]{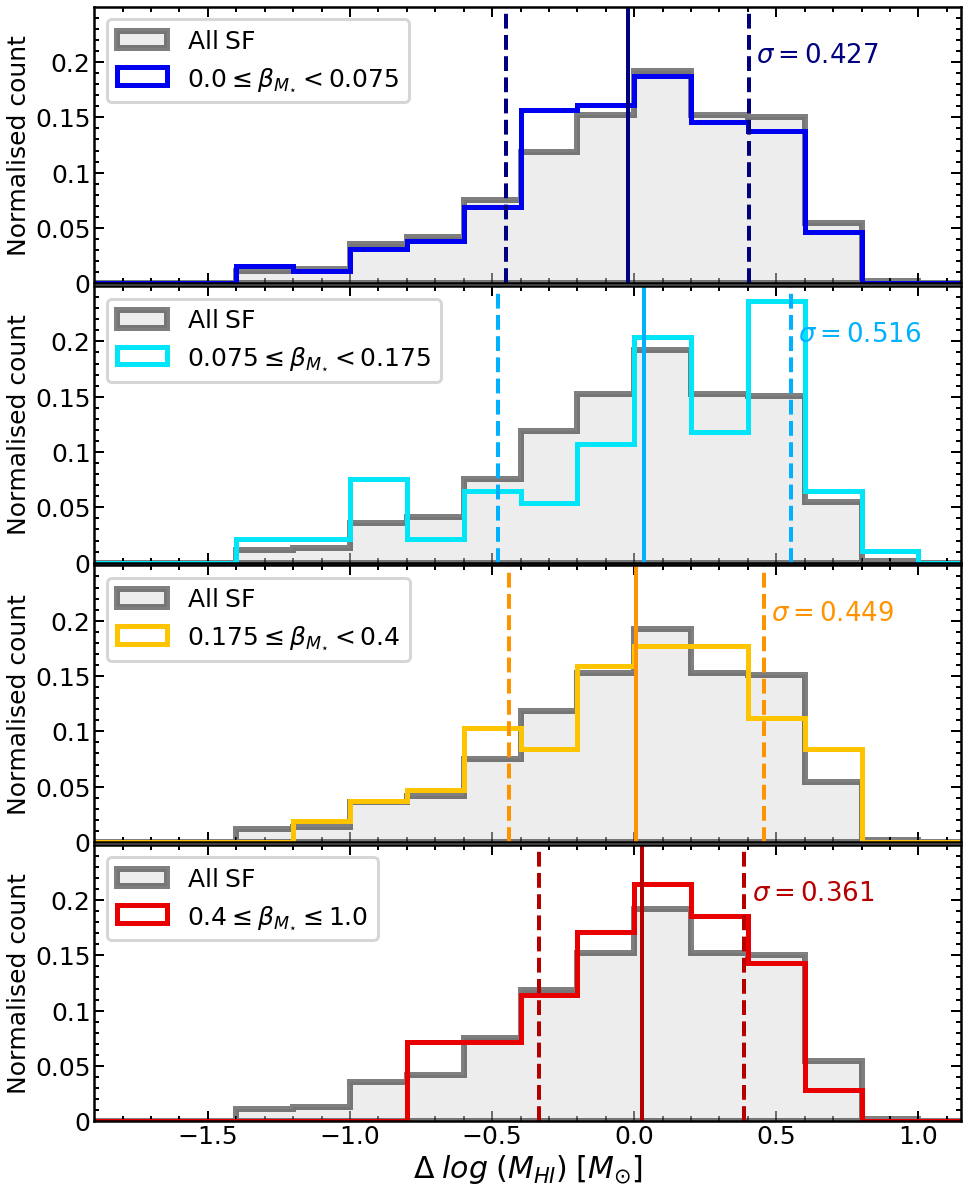}
    \caption{The distribution of $\Delta\,\log{(M_\mathrm{HI})}$ for star-forming galaxies in each of the stellar mass bulge-to-total bins shown in Figure \ref{fig:MHI_vs_Mstar_by_B2T_SFonly}. The recurring grey histogram shows the distribution for all star-forming galaxies. Solid and dashed vertical lines show the means and $\pm 1\sigma$ standard deviations, respectively. The areas under the histograms have been normalised to unity.}
    \label{fig:delta_MHI_by_B2T}
\end{figure}

In Figure \ref{fig:delta_MHI_by_B2T}, we show the distributions of \MHI for each of the bulge-to-total ratio bins shown in Figure \ref{fig:MHI_vs_Mstar_by_B2T_SFonly}. The $\Delta\,\log{(M_\mathrm{HI})}$ quantity shown is derived by first fitting a line to the \MHIvM relation shown in Figure \ref{fig:MHI_vs_Mstar_by_B2T_SFonly}, then taking the difference between the measured \MHI and that of this best fit relation. The standard deviations of these distributions range from $\sigma = 0.36\;\mathrm{dex}$ in the most bulge-dominated bin to $\sigma = 0.52\;\mathrm{dex}$. Overall, there is not a significant difference between the scatters in each bin for star-forming galaxies. After performing two-sample Kolmogorov-Smirnov tests \citep{Massey1951} between each \BtoTM bin and the complete star-forming subset, we do not find strong evidence to reject the null hypothesis that the subsets are drawn from the same parent distribution.

It is possible that any trend with bulge-to-total ratio present is not detected given the uncertainties in the measurements. The separations between medians in Figure \ref{fig:MHI_vs_Mstar_by_B2T_SFonly} are within their standard errors and, at most, differ by 0.3 dex. This equates to a maximum factor of 2 between different bins of \BtoTM suggesting that if the presence of bulges had an effect on galaxy cold gas reservoirs, their role in quenching galaxies is likely minor. In other words, the cold gas reservoirs of two star-forming galaxies with similar stellar masses in their discs do not appear significantly influenced by the presence of a central bulge.

\section{Discussion}
\label{sec:discussion}
Many previous studies using both observations and simulations have shown that the quenching of star formation in galaxies is closely correlated with their internal structure \citep{Strateva2001,Kauffmann2003,Wuyts2011,Cheung2012}. These strong correlations have often been used to argue that internal structure must naturally play an important role in governing star formation. However, it has yet to be shown that this link is due to a direct causality and, moreover, in which direction the causality might imply. In addition to this, it is not yet clear whether a similar connection exists between the internal structure of galaxies and their cold gas reservoirs. Understanding the extent of this interplay should provide useful constraints on the possible mechanisms responsible for the observed correlations with galaxy quenching.

Many observational and simulation studies have provided evidence for a direct link to the presence of a bulge through the removal or stabilisation of a galaxy's cold gas. Early simulations of galaxies by \citet{Ostriker+Peebles1973} showed that a gaseous disc could be stabilised against fragmentation if embedded within an additional spheroidal component. This has been used to explain how some bulge-dominated systems are observed to have lower star formation efficiencies than galaxies with comparable gas discs that lack a significant bulge component \citep{Saintonge2012,Martig2013}. For example, \citet{Martig2009} proposed a scenario of `morphological quenching' as a possible explanation for the existence of bulge-dominated systems which appear resistant to star formation even in the presence of a cold gas reservoirs \citep{Grossi2009,Serra2012}. They state that the star formation in a gaseous disc could be severely suppressed once embedded within a large spheroid where star-forming clumps are no longer able to balance the increased disruptive tidal forces.

In contrast, our results suggest that the present day cold gas reservoirs of star-forming galaxies are not heavily influenced by the presence of massive central bulges. As the atomic gas of \xGASS galaxies appears mostly unconcerned with internal galaxy structure, this implies that the process(es) responsible for the shut down of star formation are likely to occur at the source of inflowing cold gas rather than at the interface at which cold gas collapses into dense regions. For instance, in a scenario where morphological quenching were at play, at fixed disc stellar mass, one should expect to find that more bulge-dominated systems are systematically more gas-rich as the consumption of this cold gas should have been suppressed (i.e. a lower star formation efficiency). Figure \ref{fig:MHI_vs_Mstar_by_B2T} shows that the converse is true; bulge-dominated galaxies are, in general, less gas-rich at fixed disc stellar masses. This result is consistent with previous observational studies \citep{Schiminovich2010, Fabello2011} as well as results from studies of the cold ISM component of galaxies simulated in semi-analytic models \citep{Lagos2014}. From a homogenised compilation of local galaxy samples (including \xGASS galaxies) with morphological classifications and stellar, \HI and/or \Htwo masses, \citet{Calette2018} found that late-type galaxies have significantly higher \HI and \Htwo gas fractions than early-type galaxies over a wide range of stellar masses.

The lack of a residual trend in the population of star-forming galaxies (Figure \ref{fig:MHI_vs_Mstar_by_B2T_SFonly}) implies that the differences in cold gas properties are likely rooted in the differing star formation properties of galaxies with different morphologies. This confirms that star formation is the most closely related property to the atomic gas reservoirs in galaxies. This is in agreement with \citet{Brown2015} who used \HI spectral stacking for a volume-limited, stellar mass selected sample intersecting the SDSS DR7, ALFALFA, and GALEX surveys. The authors show that the \HI scaling relations are primarily driven by a combination of the galaxy \sSFR bimodality and the integrated Kennicutt-Schmidt law rather than a transition in structure. \citet{Koyama2019} investigated the trends with morphology using CO($J=1$--$0$) observations for a sample of 35 green-valley galaxies separated into bulge- and disc-dominated systems. They find that the distributions of molecular gas mass and SFE of green-valley galaxies do not differ with their morphology.\\

The same conclusions are reached if, instead of the bulge-to-total ratio, we use the central mass surface density within 1 kpc as a proxy for internal structure. Conversely, if we use publicly available visual morphological classifications for our sample (e.g., \citealt{Willett2013,DominguezSanchez2018}), residual trends with morphology are still present even when controlling simultaneously by star formation and stellar mass. After careful examination, this appears to be primarily due to the well-known limitation of visual classification techniques in identifying pure bulges (e.g., ATLAS3D papers; \citealt{Emsellem2011}), highlighting the importance of less subjective quantifications of galaxy structure.

The fact that the \HI content of star-forming galaxies appears largely unconcerned with the prominence of their bulge points towards a scenario in which the growth of the disc is decoupled from that of the bulge. In other words, the disc may form from the accretion of cold gas over a time that proceeded after a central spheroid has already been established in the galaxy. This follows from the idea that galaxies grow from their inside-out \citep{Brooks2009} and is expected from the size evolution observed in star-forming galaxies \citep{Brooks2009,Williams2010}.

The observed correlations between star formation rate and quantities relating to bulge prominence (e.g. bulge mass, central velocity dispersion) has generally been interpreted as evidence for a direct connection between quenching and bulge formation \citep{Franx2008,Wake2012}. The favoured physical mechanisms that are often invoked to explain such correlations are merger-driven bulge growth and feedback from the active galactic nuclei (AGN) of accreting supermassive black holes. In the case of mergers, an increase of cold gas inflows due to gravitational instabilities may lead to an increased rate of gas consumption via the interaction induced star formation \citep{Georgakakis2000}. Galaxies with more prominent bulges typically host larger black holes with more energetic AGN \citep{vandeVoort2016}. This could lead to suppressed accretion of newly cooled gas which likely inhibits the regeneration of a cold, rotationally-supported disc \citep{Dubois2013,Lagos2014}.

While our findings may appear to be in contradiction with this picture, it is important to note that the \xGASS sample is not particularly well suited to draw strong conclusions on which mechanism(s) is responsible for quenching galaxies that are already passive as they are typically not detected in these \HI observations. Thus, our results cannot be compared at face-value with those obtained by looking at properties of passive galaxies. However, the lack of a correlation between \HI gas content and bulge-to-total ratio (within a factor $\lesssim 2$) for star-forming galaxies suggests that at $z\sim 0$, the presence of bulges in main sequence galaxies do not significantly alter the interplay between cold gas and star formation. This is in agreement with recent studies by \citet{Ellison2018,Ellison2019} who find that --- at fixed stellar mass and star formation rate --- both post-merger galaxies and those hosting AGN have similar \HI contents to a sample of control galaxies.

\section{Summary and Conclusions}
\label{sec:conclusion}
We have performed structural decompositions on 1179 galaxies in \xGASS using Bayesian two-dimensional light profile fitting with \ProFit. The addition of visually-guided filtering and re-modelling with constraints has allowed us to successfully measure physically-motivated structural parameters for 91\% of the sample, surpassing the nominal success rates of previous structural decomposition studies. We use the combination of this structural information with the observations of the atomic gas properties available in the \xGASS survey to identify whether the structure of galaxies play a role in regulating cold gas reservoirs. The key results from this analysis are as follows:
\begin{enumerate}
    \item We find that, in general, the increasing prominence of the bulge is incident with galaxies of systematically lower atomic gas mass at fixed total and disc stellar mass. This trend is most distinct when separating by the i-band flux bulge-to-total ratio rather than the stellar mass bulge-to-total ratio. This could be explained if the trend is linked to the bimodality in star formation properties of which the \iband flux is more sensitive than the stellar mass.
    \item This separation in median gas mass for different bins of bulge-to-total ratio becomes less distinct when comparing galaxies at fixed disc mass. This implies that the cold gas reservoirs of discs are more alike than galaxies taken as a whole where a massive bulge containing little gas is included.
    \item By selecting only the subset of star-forming galaxies in the sample --- thereby removing the correlation between star formation and cold gas properties --- there is little-to-no evidence for a significant residual trend with the bulge-to-total ratio of galaxies. This implies that scatter in the observed \HI scaling relations for \xGASS galaxies are more likely connected to the bimodality in star formation properties with galaxy morphology playing a far lesser role, if any.
\end{enumerate}

We do not find strong evidence that --- at $z\sim 0$ --- galaxies on the star-forming main sequence can have their gaseous discs altered by the presence of a large central bulge. This work further cautions that as there is no detectable impact on the atomic gas, many of the tight correlations observed between galaxy structure ($\Sigma_{\star}$, $R_{90}/R_{50}$, $B/T$, $\sigma_{v}$, etc.) and the quenching of star formation may not be due to a causal link and should not necessarily be assumed as evidence of the driving mechanism for the quenching of galaxies. In future work, we will use the robust structural decompositions of the \xGASS to investigate the role of a bulge on the star formation rates (\SFR) and global depletion timescales ($\tau_{dep} = M_{gas}/\dot{M_{\star}}$) in galaxies.

\section*{Acknowledgements}

Parts of this research were supported by the Australian Research Council Centre of Excellence for All Sky Astrophysics in 3 Dimensions (ASTRO 3D), through project number CE170100013.

LC is the recipient of an Australian Research Council Future Fellowship (FT180100066) funded by the Australian Government.

We are grateful to the referee for their constructive input into this work.



\bibliographystyle{mnras.bst}
\bibliography{bibfile}





\appendix

\begin{table}
	\centering
	\caption{Median \MHI gas mass for scaling relations separated into bins of \iband bulge-to-total (\BtoTi) ratio shown in the top panels of Figure \ref{fig:MHI_vs_Mstar_by_B2T}. The column $\langle x\rangle$ gives the median stellar mass within a bin and $\langle M_\mathrm{H\textsc{I}}\rangle$ gives the corresponding median \HI mass. The final bulge-to-total bin in entries for $\rm{log\,M_{\star,T}}$ explicitly includes galaxies with $\beta = 1$. The column labelled N is the number of points contributing to the bin, with the number in parentheses indicating how many are \HI non-detections.}
	\begin{tabular}{ccccl}\hline
		 $x$ & \BtoTi & $\langle x\rangle$ & $\langle M_\mathrm{H\textsc{I}}\rangle$\;[$\rm{M}_{\odot}$] & N \\ \hline
		$\rm{log}\,M_{\star T}$ & $0.0 \leq \beta_{M_{\star}} < 0.075$ & 9.21 & 9.08 $\pm$ 0.07 & 120 (15) \\
		 &  & 9.69 & 9.36 $\pm$ 0.07 & 94 (11) \\
		 &  & 10.14 & 9.54 $\pm$ 0.06 & 122 (21) \\
		 &  & 10.58 & 9.70 $\pm$ 0.11 & 32 (3) \\
		 &  & 11.08 & 9.50 $\pm$ 0.27 & 2 (1) \vspace{0.2cm} \\
		 & $0.075 \leq \beta_{M_{\star}} < 0.175$ & 9.28 & 8.74 $\pm$ 0.26 & 10 (2) \\
		 &  & 9.74 & 8.89 $\pm$ 0.16 & 34 (14) \\
		 &  & 10.22 & 9.26 $\pm$ 0.09 & 61 (14) \\
		 &  & 10.77 & 9.76 $\pm$ 0.09 & 54 (12) \\
		 &  & 11.08 & 9.72 $\pm$ 0.11 & 21 (5) \vspace{0.2cm} \\
		 & $0.175 \leq \beta_{M_{\star}} < 0.4$ & 9.29 & 8.27 $\pm$ 0.39 & 5 (1) \\
		 &  & 9.67 & 8.72 $\pm$ 0.16 & 21 (7) \\
		 &  & 10.25 & 8.75 $\pm$ 0.06 & 100 (48) \\
		 &  & 10.74 & 9.12 $\pm$ 0.05 & 139 (62) \\
		 &  & 11.12 & 9.30 $\pm$ 0.05 & 67 (50) \vspace{0.2cm} \\
		 & $0.4 \leq \beta_{M_{\star}} \leq 1.0$ & 9.01 & 8.54 $\pm$ 0.00 & 1 (0) \\
		 &  & 9.71 & 8.14 $\pm$ 0.31 & 8 (5) \\
		 &  & 10.28 & 8.74 $\pm$ 0.11 & 27 (13) \\
		 &  & 10.67 & 9.18 $\pm$ 0.12 & 26 (12) \\
		 &  & 11.17 & 9.37 $\pm$ 0.12 & 16 (8) \vspace{0.2cm} \\
		 &  &  &  & \\
		$\rm{log}\,M_{\star D}$ & $0.0 \leq \beta_{M_{\star}} < 0.075$ & 9.18 & 9.08 $\pm$ 0.07 & 102 (9) \\
		 &  & 9.51 & 9.27 $\pm$ 0.08 & 88 (13) \\
		 &  & 10.06 & 9.48 $\pm$ 0.06 & 125 (23) \\
		 &  & 10.46 & 9.72 $\pm$ 0.09 & 45 (4) \\
		 &  & 10.85 & 9.44 $\pm$ 0.18 & 10 (2) \vspace{0.2cm} \\
		 & $0.075 \leq \beta_{M_{\star}} < 0.175$ & 9.20 & 8.71 $\pm$ 0.28 & 10 (3) \\
		 &  & 9.69 & 9.13 $\pm$ 0.16 & 33 (12) \\
		 &  & 10.09 & 9.26 $\pm$ 0.10 & 53 (14) \\
		 &  & 10.54 & 9.72 $\pm$ 0.11 & 45 (9) \\
		 &  & 10.90 & 9.84 $\pm$ 0.10 & 38 (9) \vspace{0.2cm} \\
		 & $0.175 \leq \beta_{M_{\star}} < 0.4$ & 9.27 & 8.38 $\pm$ 0.25 & 11 (2) \\
		 &  & 9.71 & 9.01 $\pm$ 0.15 & 23 (5) \\
		 &  & 10.08 & 8.78 $\pm$ 0.07 & 95 (45) \\
		 &  & 10.51 & 9.07 $\pm$ 0.06 & 123 (61) \\
		 &  & 10.89 & 9.27 $\pm$ 0.05 & 76 (51) \vspace{0.2cm} \\
		 & $0.4 \leq \beta_{M_{\star}} < 1.0$ & 9.15 & 8.24 $\pm$ 0.47 & 4 (2) \\
		 &  & 9.64 & 8.71 $\pm$ 0.22 & 13 (6) \\
		 &  & 10.03 & 9.18 $\pm$ 0.11 & 34 (15) \\
		 &  & 10.44 & 9.12 $\pm$ 0.12 & 17 (9) \\
		 &  & 10.88 & 9.37 $\pm$ 0.17 & 10 (7) \vspace{0.2cm} \\
		\hline
	\end{tabular}
	\label{tab:MHI_vs_Mstar_B2T_i}
\end{table}

\begin{table}
	\centering
	\caption{Median \MHI gas mass for scaling relations separated into bins of stellar mass bulge-to-total ratio shown in the bottom panels of Figure \ref{fig:MHI_vs_Mstar_by_B2T}. Columns are as per Table \ref{tab:MHI_vs_Mstar_B2T_i}.}
	\begin{tabular}{ccccl}\hline
		 $x$ & \BtoTM & $\langle x\rangle$ & $\langle M_\mathrm{H\textsc{I}}\rangle$\;[$\rm{M}_{\odot}$] & N \\ \hline
		$\rm{log}\,M_{\star T}$ & $0.0 \leq \beta_{M_{\star}} < 0.075$ & 9.21 & 9.02 $\pm$ 0.07 & 115 (16) \\
		 &  & 9.69 & 9.29 $\pm$ 0.08 & 95 (14) \\
		 &  & 10.12 & 9.38 $\pm$ 0.06 & 104 (20) \\
		 &  & 10.57 & 9.62 $\pm$ 0.12 & 22 (3) \\
		 &  & -- & -- & 0 (0) \vspace{0.2cm} \\
		 & $0.075 \leq \beta_{M_{\star}} < 0.175$ & 9.26 & 9.34 $\pm$ 0.17 & 12 (1) \\
		 &  & 9.74 & 9.27 $\pm$ 0.19 & 26 (8) \\
		 &  & 10.22 & 9.40 $\pm$ 0.11 & 49 (10) \\
		 &  & 10.69 & 9.72 $\pm$ 0.12 & 31 (6) \\
		 &  & 11.06 & 9.76 $\pm$ 0.22 & 7 (1) \vspace{0.2cm} \\
		 & $0.175 \leq \beta_{M_{\star}} < 0.4$ & 9.33 & 8.76 $\pm$ 0.30 & 7 (1) \\
		 &  & 9.77 & 8.74 $\pm$ 0.16 & 27 (11) \\
		 &  & 10.25 & 8.76 $\pm$ 0.07 & 103 (48) \\
		 &  & 10.75 & 9.14 $\pm$ 0.06 & 125 (57) \\
		 &  & 11.12 & 9.31 $\pm$ 0.05 & 69 (47) \vspace{0.2cm} \\
		 & $0.4 \leq \beta_{M_{\star}} \leq 1.0$ & 9.20 & 8.97 $\pm$ 0.61 & 2 (0) \\
		 &  & 9.70 & 8.51 $\pm$ 0.32 & 9 (4) \\
		 &  & 10.27 & 9.08 $\pm$ 0.09 & 54 (18) \\
		 &  & 10.71 & 9.44 $\pm$ 0.08 & 73 (23) \\
		 &  & 11.13 & 9.37 $\pm$ 0.09 & 30 (16) \vspace{0.2cm} \\
		 &  &  &  & \\
		$\rm{log}\,M_{\star D}$ & $0.0 \leq \beta_{M_{\star}} < 0.075$ & 9.19 & 9.02 $\pm$ 0.07 & 97 (10) \\
		 &  & 9.53 & 9.21 $\pm$ 0.08 & 90 (15) \\
		 &  & 10.05 & 9.39 $\pm$ 0.06 & 113 (23) \\
		 &  & 10.47 & 9.57 $\pm$ 0.11 & 31 (4) \\
		 &  & 10.86 & 9.31 $\pm$ 0.14 & 5 (1) \vspace{0.2cm} \\
		 & $0.075 \leq \beta_{M_{\star}} < 0.175$ & 9.17 & 9.31 $\pm$ 0.18 & 11 (1) \\
		 &  & 9.63 & 9.27 $\pm$ 0.19 & 24 (7) \\
		 &  & 10.06 & 9.37 $\pm$ 0.12 & 41 (10) \\
		 &  & 10.46 & 9.89 $\pm$ 0.12 & 33 (5) \\
		 &  & 10.89 & 9.69 $\pm$ 0.16 & 16 (3) \vspace{0.2cm} \\
		 & $0.175 \leq \beta_{M_{\star}} < 0.4$ & 9.23 & 8.70 $\pm$ 0.25 & 11 (3) \\
		 &  & 9.72 & 8.97 $\pm$ 0.15 & 26 (9) \\
		 &  & 10.09 & 8.77 $\pm$ 0.07 & 87 (44) \\
		 &  & 10.51 & 9.07 $\pm$ 0.06 & 109 (51) \\
		 &  & 10.90 & 9.31 $\pm$ 0.05 & 92 (53) \vspace{0.2cm} \\
		 & $0.4 \leq \beta_{M_{\star}} < 1.0$ & 9.12 & 8.96 $\pm$ 0.34 & 8 (2) \\
		 &  & 9.66 & 9.23 $\pm$ 0.19 & 17 (5) \\
		 &  & 10.06 & 9.37 $\pm$ 0.08 & 66 (20) \\
		 &  & 10.51 & 9.26 $\pm$ 0.08 & 57 (23) \\
		 &  & 10.86 & 9.39 $\pm$ 0.12 & 21 (12) \vspace{0.2cm} \\
		\hline
	\end{tabular}
	\label{tab:MHI_vs_Mstar_B2T_M}
\end{table}

\begin{table}
	\centering
	\caption{Median \MHI gas mass for scaling relations separated into bins of stellar mass bulge-to-total ratio for star-forming galaxies only shown in Figure \ref{fig:MHI_vs_Mstar_by_B2T_SFonly}. Columns are as per Table \ref{tab:MHI_vs_Mstar_B2T_i}.}
	\begin{tabular}{ccccl}\hline
		 $x$ & \BtoTM & $\langle x\rangle$ & $\langle M_\mathrm{H\textsc{I}}\rangle$\;[$\rm{M}_{\odot}$] & N \\ \hline
		$\rm{log}\,M_{\star T}$ & $0.0 \leq \beta_{M_{\star}} < 0.075$ & 9.20 & 9.16 $\pm$ 0.06 & 74 (1) \\
		 &  & 9.67 & 9.41 $\pm$ 0.06 & 75 (3) \\
		 &  & 10.07 & 9.49 $\pm$ 0.06 & 89 (6) \\
		 &  & 10.53 & 9.77 $\pm$ 0.10 & 20 (0) \\
		 &  & 10.94 & 9.50 $\pm$ 0.27 & 2 (0) \vspace{0.2cm} \\
		 & $0.075 \leq \beta_{M_{\star}} < 0.175$ & 9.22 & 9.31 $\pm$ 0.12 & 9 (0) \\
		 &  & 9.67 & 9.70 $\pm$ 0.18 & 17 (1) \\
		 &  & 10.12 & 9.61 $\pm$ 0.12 & 28 (2) \\
		 &  & 10.51 & 9.92 $\pm$ 0.13 & 28 (2) \\
		 &  & 11.01 & 9.95 $\pm$ 0.17 & 11 (1) \vspace{0.2cm} \\
		 & $0.175 \leq \beta_{M_{\star}} < 0.4$ & 9.22 & 8.76 $\pm$ 0.30 & 5 (0) \\
		 &  & 9.74 & 9.45 $\pm$ 0.12 & 8 (0) \\
		 &  & 10.11 & 9.58 $\pm$ 0.11 & 34 (3) \\
		 &  & 10.62 & 9.85 $\pm$ 0.10 & 33 (0) \\
		 &  & 10.96 & 9.98 $\pm$ 0.09 & 27 (4) \vspace{0.2cm} \\
		 & $0.4 \leq \beta_{M_{\star}} \leq 1.0$ & 9.25 & 9.51 $\pm$ 0.14 & 2 (0) \\
		 &  & 9.52 & 9.74 $\pm$ 0.11 & 4 (0) \\
		 &  & 10.20 & 9.53 $\pm$ 0.11 & 15 (0) \\
		 &  & 10.56 & 9.74 $\pm$ 0.08 & 33 (0) \\
		 &  & 10.96 & 9.83 $\pm$ 0.11 & 15 (2) \vspace{0.2cm} \\
		 &  &  &  & \\
		$\rm{log}\,M_{\star D}$ & $0.0 \leq \beta_{M_{\star}} < 0.075$ & 9.19 & 9.16 $\pm$ 0.06 & 70 (1) \\
		 &  & 9.58 & 9.36 $\pm$ 0.06 & 68 (2) \\
		 &  & 10.04 & 9.53 $\pm$ 0.06 & 94 (7) \\
		 &  & 10.49 & 9.72 $\pm$ 0.09 & 25 (0) \\
		 &  & 10.94 & 9.54 $\pm$ 0.18 & 3 (0) \vspace{0.2cm} \\
		 & $0.075 \leq \beta_{M_{\star}} < 0.175$ & 9.17 & 9.34 $\pm$ 0.12 & 10 (0) \\
		 &  & 9.63 & 9.58 $\pm$ 0.18 & 16 (1) \\
		 &  & 10.06 & 9.61 $\pm$ 0.12 & 28 (2) \\
		 &  & 10.43 & 9.93 $\pm$ 0.13 & 27 (2) \\
		 &  & 10.91 & 9.88 $\pm$ 0.17 & 12 (1) \vspace{0.2cm} \\
		 & $0.175 \leq \beta_{M_{\star}} < 0.4$ & 9.20 & 9.62 $\pm$ 0.27 & 5 (0) \\
		 &  & 9.72 & 9.42 $\pm$ 0.16 & 12 (1) \\
		 &  & 10.05 & 9.59 $\pm$ 0.10 & 33 (2) \\
		 &  & 10.57 & 9.85 $\pm$ 0.11 & 33 (1) \\
		 &  & 10.85 & 9.98 $\pm$ 0.10 & 23 (3) \vspace{0.2cm} \\
		 & $0.4 \leq \beta_{M_{\star}} < 1.0$ & 9.12 & 9.68 $\pm$ 0.09 & 4 (0) \\
		 &  & 9.67 & 9.71 $\pm$ 0.13 & 7 (0) \\
		 &  & 10.13 & 9.60 $\pm$ 0.09 & 31 (1) \\
		 &  & 10.48 & 9.83 $\pm$ 0.09 & 21 (0) \\
		 &  & 10.90 & 9.95 $\pm$ 0.27 & 4 (1) \vspace{0.2cm} \\
		\hline
	\end{tabular}
	\label{tab:MHI_vs_Mstar_B2T_M_SF}
\end{table}

\section{Cold Gas Scaling Relations Medians}
\label{appendix:MHI_vs_Mstar_tables}
To supplement Figures \ref{fig:MHI_vs_Mstar_by_B2T} and \ref{fig:MHI_vs_Mstar_by_B2T_SFonly}, below are the corresponding median \MHI quantities in bins of stellar mass and bulge-to-total ratio as well as the corresponding number of galaxies contributing to each bin.

\section{Segmentation Map ProFound Parameters}
\label{appendix:profound_parameters}
We perform our galaxy segmentation process utilising many of the functions available within the \ProFound \textsc{R} package \citep{Robotham2018}. See below for the basic parameters used in order to run our initial source extraction on SDSS \rband images. The parameters that have been set are as follows:
\begin{itemize}
    \item \textbf{\emph{skycut} = 1.5}; lowest threshold made for an object in units of the sky RMS.
    \item \textbf{\emph{pixcut} = 3}; minimum number of pixels required to identify an object.
    \item \textbf{\emph{tolerance} = 3.5}; minimum difference (in units of sky RMS) between highest point and the point of contact with another segment in order for it to not be combined.
    \item \textbf{\emph{ext} = 1.0}; radius from the main segment for the detection of neighbouring objects.
    \item \textbf{\emph{sigma} = 1.5}; standard deviation of the blur used when smoothing the image.
    \item \textbf{\emph{size} = 51}; the size (in pixels) of the dilation kernel.
\end{itemize}

\section{Cold Gas Scaling Relations using Previous Literature}
\label{appendix:MHI_vs_Mstar_S11_and_M15}

\begin{figure*}
    \centering
    \includegraphics[width=\textwidth]{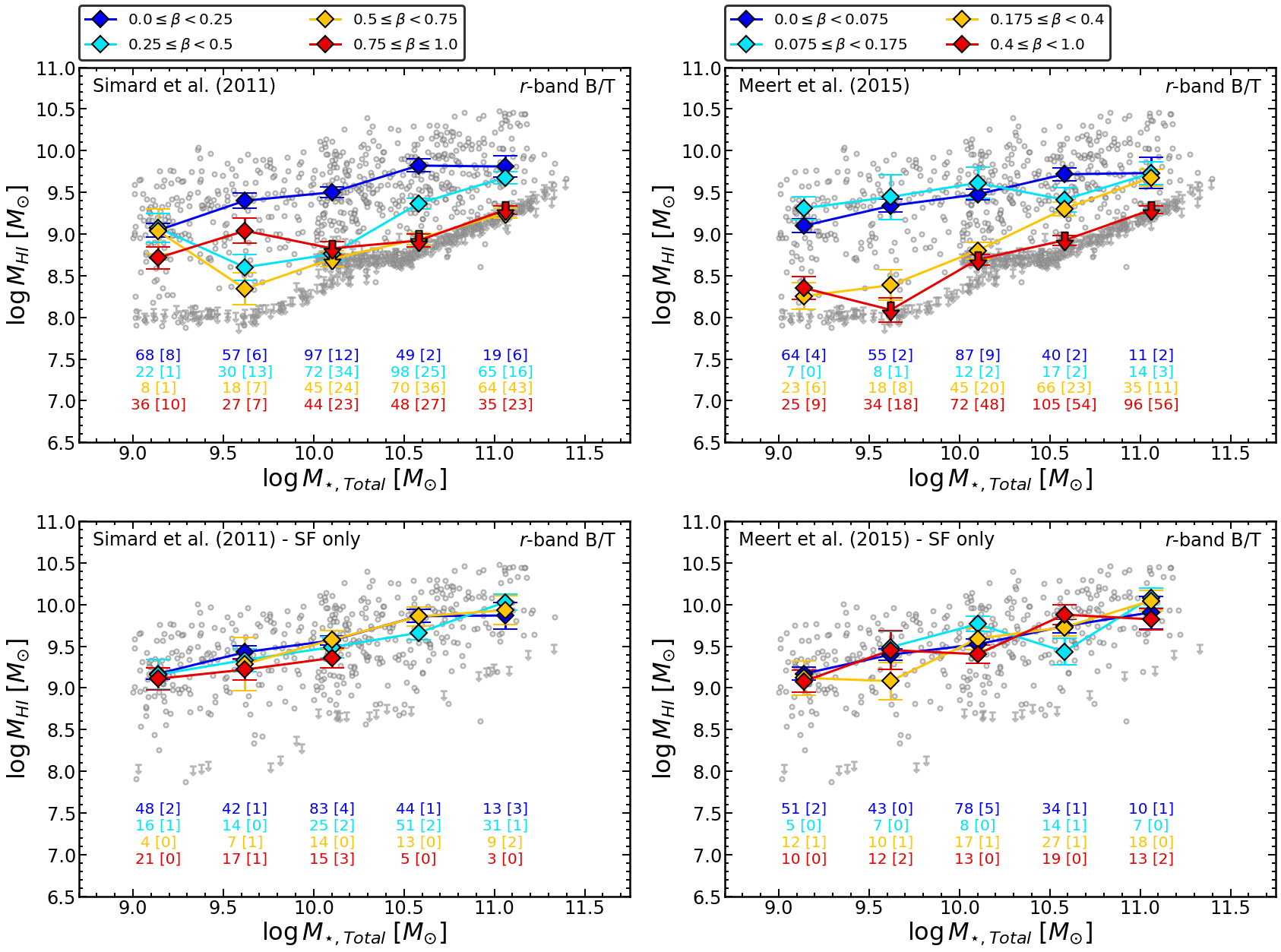}
    \caption{The equivalent relations as Figures \ref{fig:MHI_vs_Mstar_by_B2T} and \ref{fig:MHI_vs_Mstar_by_B2T_SFonly} but instead using \rband bulge-to-total ratios from the \citealt{Simard2011} (left panels) and \citealt{Meert2015} (right panels) catalogues. In the top panels we show the full \xGASS sample and in the bottom panels we show only the subset of star-forming galaxies, as defined in section \ref{sec:role_of_bulges_in_SFGs}. Note that we have had to adjust the intervals of \BtoTr for the \citet{Simard2011} catalogue to reflect the different range of values measured (see Figure \ref{fig:literature_B2T_comparisons}).}
    \label{fig:MHI_vs_Mstar_by_B2T_literature}
\end{figure*}

Here, we show Figures \ref{fig:MHI_vs_Mstar_by_B2T} and \ref{fig:MHI_vs_Mstar_by_B2T_SFonly} obtained using bulge-disc decomposition catalogues from \citet{Simard2011} and \citet{Meert2015}. Note that we limit this comparison to the photometric bulge-to-total ratios and that we adjust the intervals of plots based on \citet{Simard2011} decompositions to reflect the different distribution of \BtoTr (see Figure \ref{fig:literature_B2T_comparisons}). While the overall trends appear similar between the catalogues, there are important differences. In particular, plots based on the \citet{Simard2011} decompositions show an unexpected abundance of low mass, bulge-dominated galaxies that are star-forming, which do not resemble such a morphology from visual inspection. This stems from the difficulties in selecting pure-bulge galaxies from the model residuals and a cut in \Sersic index alone. Qualitatively, our results agree more closely with the \citet{Meert2015} catalogue, albeit the separation between medians is less clear-cut when all galaxies are included.\\

\begin{figure*}
    \centering
    \includegraphics[width=0.49\textwidth]{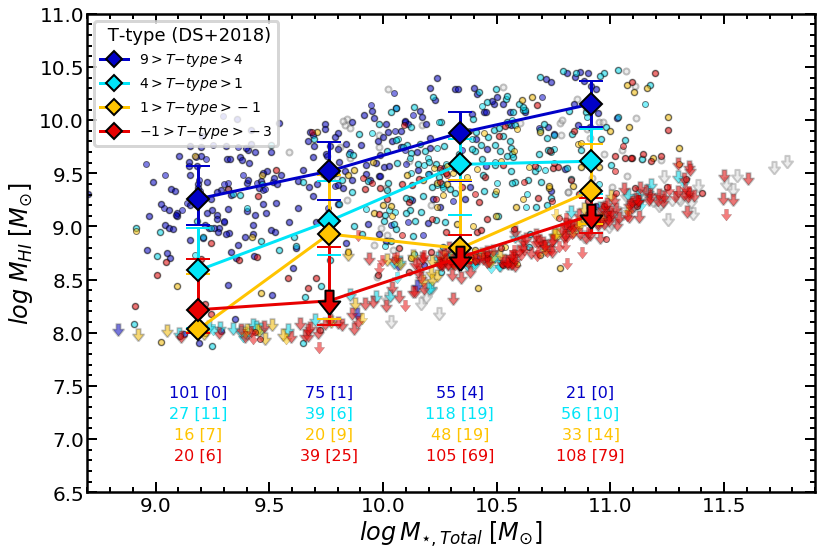}
    \includegraphics[width=0.49\textwidth]{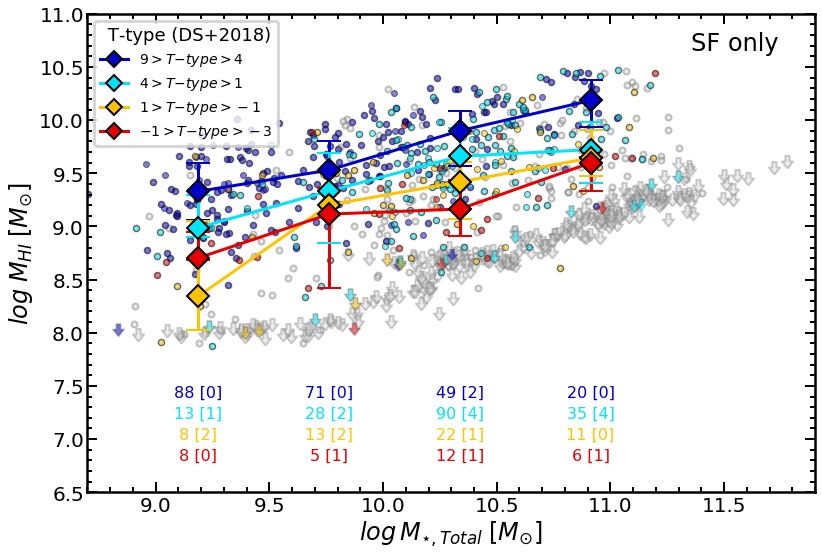}
    \includegraphics[width=0.925\textwidth]{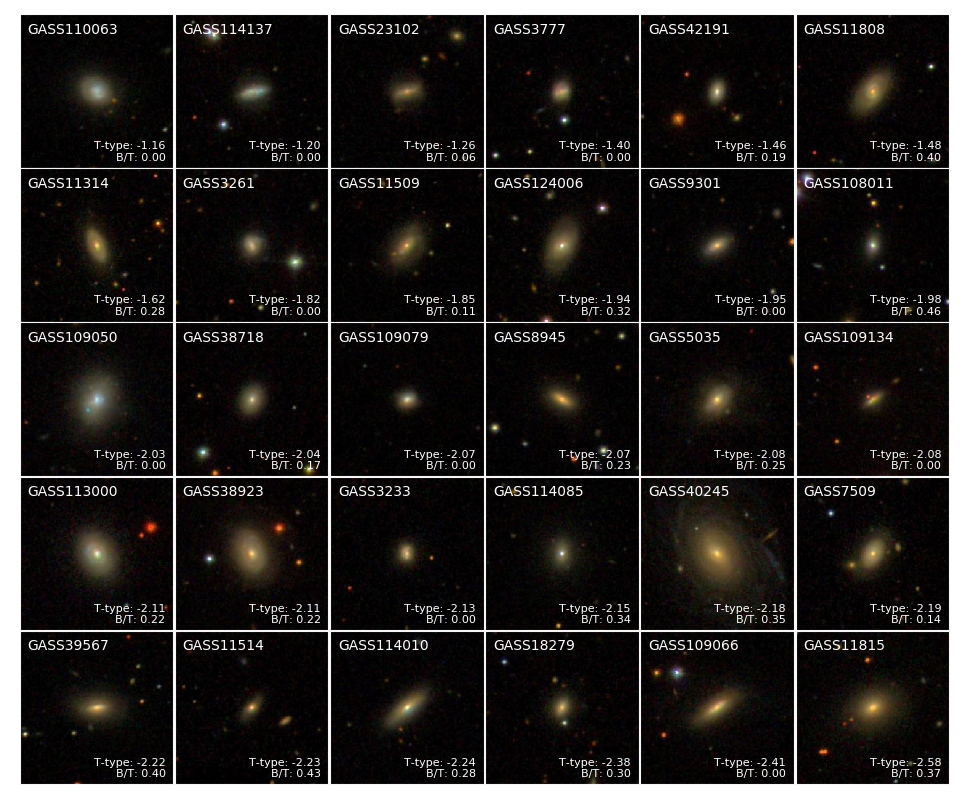}
    \caption{Top panels show the equivalent relations as Figures \ref{fig:MHI_vs_Mstar_by_B2T} and \ref{fig:MHI_vs_Mstar_by_B2T_SFonly} but instead using morphological T-types derived from \citet{DominguezSanchez2018}. The bottom panel shows the \SDSS RGB cutout images corresponding to the subset of star-forming galaxies with T-type $< -1$ (red points in top-right panel) and for which we measure a \BtoTi $< 0.5$ with photometric bulge-disc decomposition. The relevant morphological quantities are labelled in each panel.}
    \label{fig:MHI_vs_Mstar_by_Ttype}
\end{figure*}

Finally, for completeness, we investigate the trends with cold gas mass observed when visually classified morphological T-types are used. In this case, we utilise a catalogue of accurate T-types measured in \citet{DominguezSanchez2018} obtained using convolutional neural networks in combination
with existing visual classification catalogues from \citet{Nair+Abraham2010}. The left panel of Figure \ref{fig:MHI_vs_Mstar_by_Ttype} confirms our previous result that the gas reservoirs of galaxies at all star-formation rates is separated in morphology. However, when controlling for star-forming galaxies only (right panel), we see that a residual trend remains with T-type whereas this is not observed from bulge-disc decomposition models. Below, we show the \SDSS RGB cutout images of all star-forming galaxies with T-type < -1 and i-band $B/T < 0.5$; i.e. where these morphological catalogues disagree most. Many of these systems show disc-like structures (albeit not always blue), indicating that a comparison between morphological classifications and modelled bulge-to-total ratios is not necessarily straightforward.

\bsp	
\label{lastpage}
\end{document}